\documentclass[ reprint, superscriptaddress, amsmath,amssymb,
aps, pra, showkeys ]{revtex4-2}

\usepackage{soul}
\usepackage{float}
\usepackage{multibib}
\usepackage[linktocpage]{hyperref}
\usepackage{xcolor}
\hypersetup{
    colorlinks,
    linkcolor={blue!50!blue},
    citecolor={blue!50!blue},
    urlcolor={blue!80!black}
}
\usepackage{siunitx}
\usepackage{import}
\usepackage{placeins}
\usepackage{xcolor}
\usepackage{lmodern}
\usepackage[braket, qm]{qcircuit}
\usepackage{graphicx}
\usepackage{dcolumn}
\usepackage{bm}
\usepackage{comment}
\usepackage[english]{babel}
\usepackage{blindtext}
\usepackage{multirow}
\usepackage{array}
\usepackage{enumerate}
\let\oldaddcontentsline\addcontentsline
\newcommand{\stoptocentries}{\renewcommand{\addcontentsline}[3]{}}
\newcommand{\starttocentries}{\let\addcontentsline\oldaddcontentsline}

\begin{document}
\title{Quantum {{Optics}} Applications of Hexagonal Boron Nitride {{Defects}}}  

\author{Asl{\i} \surname{\c{C}akan}}
\affiliation{Department of Computer Engineering, TUM School of Computation, Information and Technology, Technical University of Munich, 80333 Munich, Germany}
\affiliation{Munich Center for Quantum Science and Technology (MCQST), 80799 Munich, Germany}

\author{Chanaprom Cholsuk}
\affiliation{Department of Computer Engineering, TUM School of Computation, Information and Technology, Technical University of Munich, 80333 Munich, Germany}
\affiliation{Munich Center for Quantum Science and Technology (MCQST), 80799 Munich, Germany}
\affiliation{Abbe Center of Photonics, Institute of Applied Physics, Friedrich Schiller University Jena, 07745 Jena, Germany}

\author{Angus Gale}
\affiliation{School of Mathematical and Physical Sciences, Faculty of Science, University of Technology Sydney, Ultimo, New South Wales 2007, Australia}
\affiliation{ARC Centre of Excellence for Transformative Meta-Optical Systems (TMOS), University of Technology Sydney, Ultimo, New South Wales 2007, Australia}

\author{Mehran Kianinia}
\affiliation{School of Mathematical and Physical Sciences, Faculty of Science, University of Technology Sydney, Ultimo, New South Wales 2007, Australia}
\affiliation{ARC Centre of Excellence for Transformative Meta-Optical Systems (TMOS), University of Technology Sydney, Ultimo, New South Wales 2007, Australia}

\author{Serkan Pa\c{c}al}
\affiliation{Department of Physics, Izmir Institute of Technology, 35430 Izmir, Turkey}

\author{Serkan Ate\c{s}}
\affiliation{Department of Physics, Izmir Institute of Technology, 35430 Izmir, Turkey}

\author{Igor Aharonovich}
\affiliation{School of Mathematical and Physical Sciences, Faculty of Science, University of Technology Sydney, Ultimo, New South Wales 2007, Australia}
\affiliation{ARC Centre of Excellence for Transformative Meta-Optical Systems (TMOS), University of Technology Sydney, Ultimo, New South Wales 2007, Australia}

\author{Milos Toth}
\affiliation{School of Mathematical and Physical Sciences, Faculty of Science, University of Technology Sydney, Ultimo, New South Wales 2007, Australia}
\affiliation{ARC Centre of Excellence for Transformative Meta-Optical Systems (TMOS), University of Technology Sydney, Ultimo, New South Wales 2007, Australia}

\author{Tobias Vogl}%
\email{tobias.vogl@tum.de}
\affiliation{Department of Computer Engineering, TUM School of Computation, Information and Technology, Technical University of Munich, 80333 Munich, Germany}
\affiliation{Munich Center for Quantum Science and Technology (MCQST), 80799 Munich, Germany}
\affiliation{Abbe Center of Photonics, Institute of Applied Physics, Friedrich Schiller University Jena, 07745 Jena, Germany}


\begin{abstract}
Hexagonal boron nitride (hBN) has emerged as a compelling platform for both classical and quantum technologies. In particular, the past decade has witnessed a surge of novel ideas and developments, which may be overwhelming for newcomers to the field. This review provides an overview of the fundamental concepts and key applications of hBN, including quantum sensing, quantum key distribution, quantum computing, and quantum memory. Additionally, we highlight critical experimental and theoretical advances that have expanded the capabilities of hBN, in a cohesive and accessible manner. The objective is to equip readers with a comprehensive understanding of the diverse applications of hBN, and provide insights into ongoing research efforts.
\end{abstract}


\maketitle
\tableofcontents
\pagenumbering{arabic}

\section{Introduction}
Hexagonal boron nitride (hBN) has attracted considerable interest recently due to its distinctive properties and potential uses in quantum technologies. This two-dimensional (2D) material, similar in structure to graphene, consists of alternating boron and nitrogen atoms arranged in a hexagonal lattice.
Its exceptional properties, such as a wide bandgap ($\sim$ 6 eV), and high thermal conductivity, chemical stability and mechanical strength, make it appealing for a wide range of applications.\\
\indent
Synthesis and initial investigations of boron nitride date back to the early 20th century. The first successful synthesis of hBN was reported by Balmain in 1842~\cite{Balmain:1842}, although its structure and properties were not well understood until much later. In the 1950s, improvements in X-ray diffraction and electron microscopy techniques enabled a more thorough exploration of the hBN crystal structure, uncovering its layered hexagonal arrangement similar to that of graphite~\cite{Pease:1950}.\\
\indent
The rise of 2D materials began with the isolation of graphene in 2004 by Geim and Novoselov~\cite{Novoselov:2004}, who were awarded the Nobel Prize in Physics in 2010 for their work. This discovery triggered growing interest in other 2D materials, including hBN. Researchers quickly realized that hBN is an ideal substrate for 2D electronic devices based on graphene.
Techniques for isolating hBN monolayers, similar to the mechanical exfoliation method used for graphene, were refined in the mid-2000s~\cite{Novoselov:2005, Dean2010-hw}. This development marked a significant milestone, enabling researchers to study the properties of hBN at the monolayer and few-layer scales. The ability to create high-quality, atomically thin hBN sheets was crucial for investigating its interactions with other 2D materials and for its use as a substrate in electronic and optoelectronic devices.\\
\indent
Explorations of hBN for quantum applications gained momentum in 2016, when it was observed to host quantum emitters exhibiting bright and stable single-photon emission at room temperature~\cite{Tran:2016a, Aharonovich:2016}. These quantum emitters, often associated with defect centers in the hBN lattice, opened new avenues for research in quantum optics and quantum information processing.
The identification and characterization of quantum emitters in hBN was the result of extensive research into defects and impurities in the material. Early research into the photoluminescence of hBN revealed the presence of deep ultraviolet emissions, hinting at the potential for UV devices~\cite{Watanabe:2004}. The discovery of room-temperature quantum emitters~\cite{Tran:2016a} was part of a broader effort to understand and control defect states in hBN, leveraging techniques from materials science and quantum optics.\\
\indent
In parallel, foundational studies revealed hBN’s distinctive hyperbolic properties, demonstrating its potential for intense light confinement and high-quality resonances~\cite{Caldwell:2014, Low:2017}. These findings established hBN as a natural hyperbolic material capable of supporting deeply sub-diffractional confinement of polaritons, a key requirement for manipulating light at the nanoscale. Ref.~\cite{Caldwell:2014} was pivotal since it was among the first to experimentally demonstrate the potential of hBN for hosting high-quality factor resonances, thus enabling strong light-matter interactions. Building on this, research in Ref.~\cite{Low:2017} further highlighted the integration of hBN into practical nanophotonic devices, reinforcing its role in the development of quantum technologies such as ultra-sensitive infrared sensors and components for quantum information processing.\\
\indent
The study of Van der Waals heterostructures, where hBN is stacked with other 2D materials, represents a significant chapter in the history of hBN research. Pioneering work in this area~\cite{Cao:2018} has demonstrated the ability to create heterostructures with tailored electronic and optical properties, enabling novel quantum phenomena such as Moiré superlattices and the exploration of strongly correlated electronic states. This line of research draws from a rich history of semiconductor physics and nanotechnology, building on decades of advances in material synthesis and characterization~\cite{Geim:2013}.\\
\indent
Recent advances have further expanded the understanding and utility of hBN in quantum applications. While the identification of spin defects, similar to the well-known nitrogen-vacancy center in diamond, has highlighted the potential of hBN for quantum sensing~\cite{Gottscholl:2020}, its integration with other 2D materials, such as transition metal dichalcogenides, has led to the development of novel Van der Waals heterostructures with tunable quantum properties~\cite{Liu:2016}. In addition, advances in nanofabrication and material characterization techniques, such as electron microscopy and atomic force microscopy, have further propelled studies of hBN, enabling defect visualization and manipulation at the atomic scale, and providing unprecedented insights into the material's quantum properties~\cite{Geim:2013}.\\
\indent 
The unique combination of layered and quantum properties of hBN and its compatibility with existing semiconductor technologies hold promise for scalable and practical quantum devices. Ongoing explorations of defect engineering and heterostructure fabrication techniques aim to unlock new functionalities, paving the way for next-generation quantum technologies, which will be discussed in detail in the following sections~\ref{sec:basicconcepts} and~\ref{sec:main}.\\
\indent
\section{Basic Concepts} \label{sec:basicconcepts}
Before diving into the details of applications and techniques, this section will briefly introduce the basic concepts behind the main discussions in this paper. These include hBN defects, fabrication strategies and characterization properties.

\subsection{hBN Defects} 
\begin{figure*}
    \includegraphics[width=.99\textwidth]{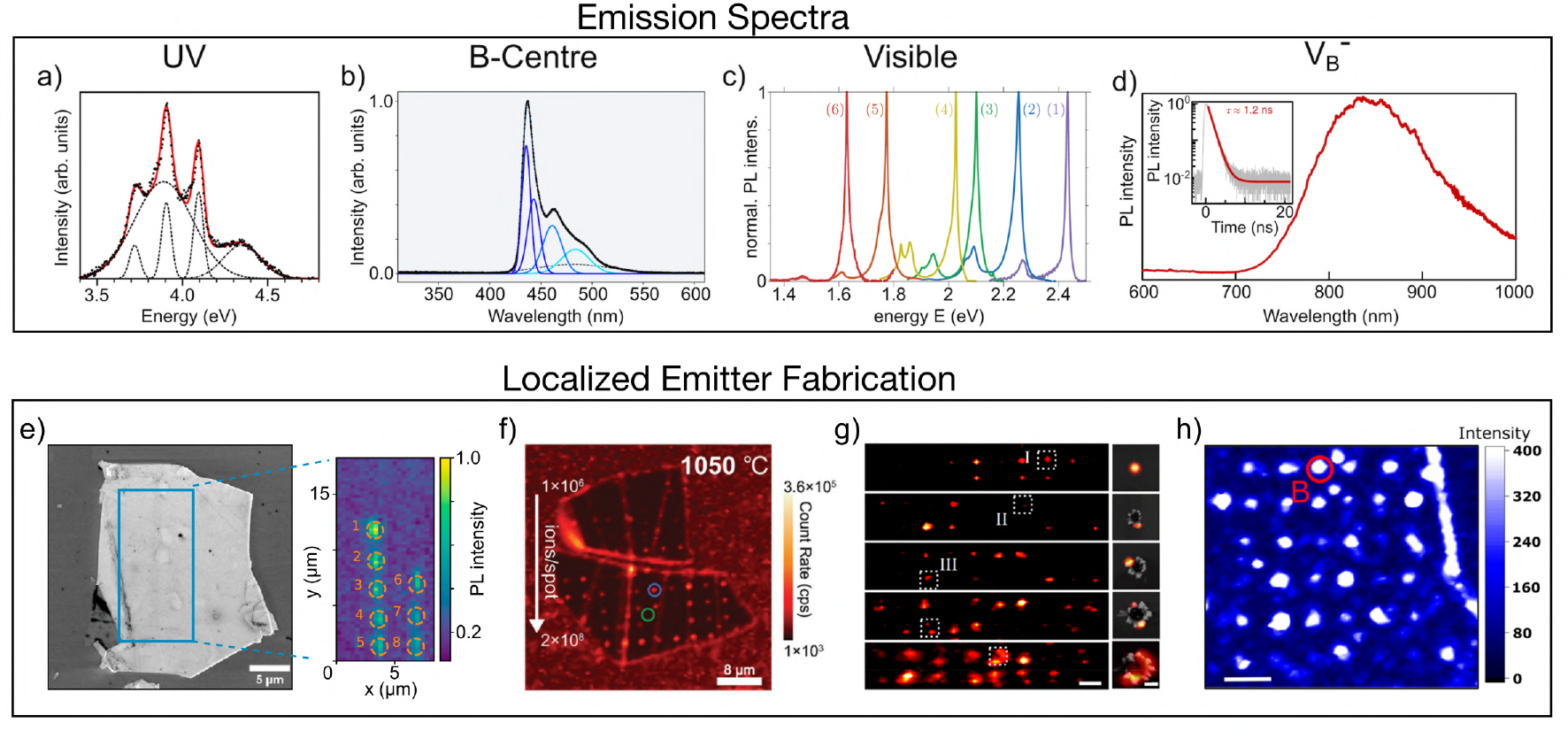} 
	\caption{ Overview of hBN defect classes and defect engineering methods. (a) Room temperature CL spectrum of the UV-defect with a ZPL at 4.1 eV and two prominent PSB emission peaks~\cite{Bourrellier:2016}. This figure is adapted with permission from Ref.~\cite{Bourrellier:2016}. Copyright 2016 American Chemical Society. (b) Room temperature CL spectrum of the B-centre emitter~\cite{Shevitski:2019}. This figure is reprinted with permission from Ref.~\cite{Shevitski:2019}. Copyright (2019) by the American Physical Society.  (c) PL spectra from visible single photon emitters with ZPL emissions in the range of $\sim$1.6 to $\sim$2.4 eV~\cite{Wigger:2019}. This figure is adapted from Ref.~\cite{Wigger:2019} available under a CC-BY 3.0 license. Copyright 2019 Wigger \textit{et al.} (d) PL spectrum of the V$_\text{B}^{-1}$ defect. (Inset) Decay curve of the V$_\text{B}^{-1}$ showing a lifetime of 1.2 ns~\cite{Gottscholl:2020}. This figure together with the inset is adapted from Ref.~\cite{Gottscholl:2020} with permission from Springer Nature. (e) B-centre ensembles generated by electron beam irradiation. Left: Scanning electron micrograph of an hBN flake. Right: PL map corresponding to the blue outlined region in the micrograph~\cite{Fournier:2021}. The figures are adapted from Ref.~\cite{Fournier:2021} available under a CC-BY 4.0 license. Copyright 2021 Fournier \textit{et al.} (f) PL map of emitters engineered by He+ ion beam irradiation followed by annealing in an oxygen atmosphere~\cite{Liu:2023}. This figure is adapted from Ref.~\cite{Liu:2023} available under a CC-BY-NC 4.0 license. Copyright 2023 Liu \textit{et al.} (g) Bright localised PL emissions generated by femtosecond laser irradiation (scale bars = 2 µm)~\cite{Gan:2022}. This figure is adapted with permission from Ref.~\cite{Gan:2022}. Copyright 2022 American Chemical Society. (h) PL map showing single photon emitters engineered by atomic force microscopy indentation (scale bar = 2 µm)~\cite{Xu:2021}. This figure is adapted with permission from Ref.~\cite{Xu:2021}. Copyright 2021 American Chemical Society.
}
\label{Fig:secII}
\end{figure*}

In recent years, a number of studies have explored optically-active defects (i.e., color centres) in layered Van der Waals materials. These materials are unique when compared to their bulk counterparts, due to their ability to be exfoliated and assembled into heterostructures. Most color centres in layered materials studied to date are hosted by hBN. Defects in hBN are varied in their structure and properties, with emissions spanning the Ultraviolet (UV) to Near-Infrared (NIR) spectral range. They can be classed broadly by emission wavelength into the following four groups. 
\begin{itemize}
    \item UV emitters with a zero phonon line (ZPL) at 4.1 eV (305 nm). These emitters have a distinct spectral shape with the two prominent phonon side band (PSB) emissions seen in Figure \ref{Fig:secII}(a). Although generally studied as ensembles, single defects have been identified in cathodoluminescence (CL) studies~\cite{Bourrellier:2016}. 
While no definitive defect structure has been identified, it is speculated to be a carbon complex based on both experimental and theoretical data. The UV emitters are present in carbon-rich regions of hBN based on secondary ion mass spectroscopy (SIMS) measurements~\cite{Onodera:2019} as well as samples doped with carbon post-growth~\cite{Onodera:2020}.  Theoretical studies have focused on the carbon dimer C$_\text{B}$C$_\text{N}$~\cite{Mackoit-Sinkeviciene:2019, Plo:2024} and short carbon chains or rings~\cite{Li:2022} as possible candidates for the atomic structure of the defect.

\item Blue emitters or B-centres are a class of defect with a ZPL at 2.85 eV (436 nm) and two distinct PSB emission peaks separated by 160 meV (Figure~\ref{Fig:secII}(b)). B-centres were originally identified by Shevitski et. al. as CL-active color centres~\cite{Shevitski:2019}. These emitters are unique in that they can be engineered by irradiation with a kiloelectronvolt electron beam, allowing for site-specific generation of single defects with a consistent emission wavelength using a scanning electron microscope~\cite{Fournier:2021}. The emission is linearly polarized, with dipoles oriented preferentially along three directions separated by 60°~\cite{Horder:2024}. The atomic structure of the defect is also yet to be identified. A split interstitial nitrogen vacancy has been proposed as a possible configuration~\cite{Ganyecz:2024}. Notably, the ability to generate B-centres using an electron beam correlates with the presence of the UV emitter in hBN~\cite{Gale:2022}. A recent theoretical work has suggested the carbon tetramer (C$_\text{B}$C$_\text{N}$)$_2$ defect~\cite{Maciaszek:2024}, consistent with this correlation.
\item Defects emitting in the visible spectral region, at wavelengths longer than 436 nm, are the original quantum emitters discovered in hBN~\cite{Tran:2016a}. These emitters are typically bright and possess a narrow ZPL with weak phonon coupling. Unlike the UV and B-centre defects, the visible emitters exhibit a wide range of ZPL wavelengths spanning 1.6–2.4 eV (516–775 nm), as is illustrated by Figure~\ref{Fig:secII}(c)~\cite{Wigger:2019}. A number of visible emitters possess an optically-addressable electronic spin state~\cite{Stern:2022}.
As with the other defect classes outlined thus far, the defect structures are a matter of debate and carbon impurities have been proposed to play a role~\cite{Mendelson:2021, Jara:2021, Auburger:2021, Maciaszek:2022}. Similar to the B-centre, some visible defects have threefold polarisation axes separated by 60°~\cite{Zhong:2024}.

\item The negatively charged boron vacancy (V$_{\text{B}}^{-1}$) has a broad phonon-assisted emission spectrum centred on approximately 800 nm (Figure~\ref{Fig:secII}(d))~\cite{Gottscholl:2020}. The ZPL is dark due to symmetry~\cite{Libbi:2022}, though it has been proposed to be located at $\sim$773 nm using cavity-coupled ensembles~\cite{Qian:2022}. At present, the V$_{\text{B}}^{-1}$ defect is the only emitter in hBN with a confirmed atomic structure. It was the first defect in hBN shown to possess optically-addressable spin properties~\cite{Gottscholl:2020}. However, in contrast to all of the above defect classes in hBN, V$_{\text{B}}^{-1}$ emitters have to date been studied only as ensembles due to a low quantum efficiency of the symmetry-forbidden emission.
\end{itemize}
\indent 
For defects emitting in the visible spectral region, low temperature studies have reported linewidths below 1 GHz~\cite{Konthasinghe:19,Tran2017ResonantEO}. In some instances, Fourier limited linewidths have also been reported for emitters in hBN~\cite{Dietrich2018,Hoese2022}. However, spectral diffusion is prevalent for many defects in hBN, leading to significant inhomogeneous linewidth broadening. Schemes utilized to reduce spectral diffusion and attain near homogenous linewidths include optical re-pumping with a non-resonant laser~\cite{White:2020} and electrical control via heterostructure devices~\cite{Akbari:2022}. Unlike the visible light emitting defects, the consistent ZPL of the B-centre lends itself to resonant excitation measurements, with linewidths in the range of $\sim$1.5-4 GHz reported for these defects~\cite{Horder:2022, Fournier:2023}. Although these linewidths are relatively broad compared to other hBN defects, spectral hole-burning spectroscopy was utilized for B-centres and measured a homogenous linewidth of 150 MHz~\cite{Horder:2024}. Recently, two-photon interference was achieved using a single B-centre, demonstrating photon indistinguishability for the first time from a defect in hBN~\cite{Fournier:2023}.

\subsection{Defect Fabrication}
A number of methods currently exist to engineer emitters in hBN. In the ideal case, these methods would be site-specific and deterministic in nature, with an ability to generate single photon emitters on-demand with a desired emission wavelength and polarization axis. Earlier works utilized non-localised techniques to engineer defects, such as annealing~\cite{Tran:2016b, Mohajerani:2024},  chemical treatments~\cite{Chejanovsky2016-yt}, delocalized irradiation by keV and MeV electrons~\cite{Vogl2019-ns,NgocMyDuong:2018}, as well as plasma treatments~\cite{Vogl_2017,Vogl2018-do,Chen:2021}. In contrast, recent studies have emphasized site-specific techniques, with a focus on integrating single photon emitters in electronic devices and photonic nanostructures. The most common techniques are highlighted below. They are stochastic in nature, and the success rate of generating single defects is therefore Poisson-limited to $\sim$37\%.

\subsubsection{Electron Irradiation}
B-centres can be generated by electron beam irradiation using an scanning electron microscope (Figure~\ref{Fig:secII}e))~\cite{Fournier:2021}. The emitter density is controlled by electron dose~\cite{Roux:2022}, and a success rate of $\sim$33\% has been observed experimentally for the generation of individual single photon emitters~\cite{Gale:2022}. Electron beam irradiation has also been used to activate single photon emitting in the yellow spectral range. Kumar~\textit{et al.} demonstrated localised fabrication of emitters with a PL peak at $\sim$ 575 nm~\cite{Kumar:2023,doi:10.1021/acsnano.3c08940}.

\subsubsection{Ion Irradiation}
Ion irradiation is a common method for deterministic creation of defects in solids. In hBN,
focused ion beam (FIB) methods have been used to control the position and density of V$_\text{B}^{-1}$
defects~\cite{Kianinia:2020, Sarkar:2024}. Ion irradiation can also be used to engineer single defects emitting in the
visible spectral region. The combination of helium ion irradiation and subsequent thermal
annealing in oxygen was able to generate localised visible class single photon emitters with a creation yield
of $35\%$~\cite{Liu:2023}. An array of defects using this methodology are clearly observable as isolated
fluorescent spots in Figure~\ref{Fig:secII}(f).

\subsubsection{Femtosecond Laser Writing}
Femtosecond laser writing methods have been utilized to engineer V$_\text{B}^{-1}$ ensembles in hBN around ablation sites~\cite{Gao:2021b}. Laser irradiations have also been used to engineer visible class single photon emitters (Figure~\ref{Fig:secII}(g))~\cite{Hou:2018}. More recently, Gan et. al. used a combination of single pulse femtosecond laser irradiation, ozone treatment and air annealing to generate defects in hBN with optimized conditions resulting in a 43$\%$ single photon emitter yield~\cite{Gan:2022}.
\subsubsection{Atomic Force Microscopy}
Indentation via atomic force microscopy followed by annealing in argon has also been successful in generating single photon emitters~\cite{Xu:2021}. A 36\% yield of single photon emitters was achieved using indents with a lateral size of 400 nm. Bright luminescent regions are visible in Figure \ref{Fig:secII}(h) corresponding to indented sites.

\subsubsection{Thermal Annealing}
Thermal annealing is an essential technique for enhancing single-photon emitters in hBN, significantly improving emitter density, stability, and optical quality. Annealing at temperatures around 850–1100 °C in usually inert environments enhances photon purity and narrows the ZPL linewidth by removing surface contaminants~\cite{Li:2019}. Additionally, high-temperature annealing with controlled oxygen flow reduces background fluorescence and promotes a uniform distribution of emitters, making it highly effective for quantum applications~\cite{Chen:2021}.

Thermal annealing also enables tuning of emitter properties. For instance, annealing ion-implanted hBN can shift emission wavelengths by altering defect configurations, providing precise control over spectral characteristics~\cite{Venturi2024}. This process not only stabilizes defects, maintaining spectral consistency even through further annealing cycles, but also maximizes ZPL brightness when combined with oxygen plasma etching, creating emitters that are ideal for integration into photonic circuits~\cite{Tran:2016b, Vogl2018-do}.
\subsubsection{Gas Treatment Methods}
Gas treatments can significantly influence the properties of hBN, particularly by modifying its surface and defect structures~\cite{Chen:2021}. Processes such as high-temperature annealing in oxygen or hydrogen environments and plasma treatments are effective for cleaning organic residues, repairing defects, and creating new vacancies. For instance, hydrogenation plays a critical role in modifying charge states of defects, which can stabilize quantum emitters and influence their optical properties by reducing non-radiative recombination~\cite{Tran:2016b}. These methods improve the material's optical properties, including the generation of high-purity single-photon emitters with ZPLs, enhanced photostability, and reduced background fluorescence. Specifically for quantum emitters, gas treatments such as oxygen-rich annealing at optimized conditions can significantly increase the density and quality of single photon emitters, while plasma treatments, followed by annealing, refine the emitters by removing unwanted impurities~\cite{Li:2019,Chen:2021}.
\\

\indent
At this stage, only the B-centre and V$_\text{B}^{-1}$ defects can be created with deterministic engineering methods whilst minimizing damage or modification to the hBN crystal lattice. Damage limits quantum application due to charge fluctuations at charge traps. Moving forward, new methods would hope to overcome limitations seen with the creation of visible class defects, ideally achieving both localization and controlled ZPL distribution with high yields. Similarly, for the UV defect, although the ZPL emission is consistent, there are still no localized engineering methods available.
\section{Main Applications} \label{sec:main}
The variety of hBN properties that have been uncovered over the last decade and produced state-of-the-art applications is quite extensive. Thus, we focus on the general framework of hBN single photon emitters, and emphasize selected advances in the field that are crucial for cutting-edge applications. Implementation details are not explicitly shown; the interested reader is encouraged to refer to the papers cited in the respective subsections.
\subsection{Quantum Sensing} \label{subsec:qsensing}
A quantum system is usually coupled efficiently to its local environment, making it an ideal
candidate for sensing applications. Wide-range temperature detection using single photon emitters
in hBN has been proposed, as they can operate from cryogenic temperatures to as high as 500 $^\circ$C~\cite{Kianinia:2017}. However, most research on quantum sensing using hBN is motivated by the
discovery of optically-active spin defects in this material~\cite{Gottscholl:2020, Chejanovsky2016-yt, Stern:2022,Mendelson:2021}. The electron spin
states of these defects can be readily accessed using optically detected magnetic resonance (ODMR) spectroscopy, enabling
the detection of magnetic and electric fields, strain and temperature. The first room-temperature spin defects
in hBN were reported in 2020, and indexed as negatively-charged boron vacancies (V$_{\text{B}}^{-1}$), based on
detailed studies of hyperfine interactions in conjunction with ion irradiation of hBN
crystals~\cite{Gottscholl:2020, Chejanovsky2016-yt, Stern:2022,Mendelson:2021, Kianinia:2020}. The V$_{\text{B}}^{-1}$ is a spin triplet defect with a zero-field splitting of 3.5 GHz in the ground
state, emitting broad photoluminescence (PL) at 800 nm~\cite{Gao:2021b, Mathur2022-yf, Durand:2023}. Despite a short, measured lifetime of
1.2 ns, it exhibits low brightness indicative of a very low quantum efficiency. Consequently, no measurement of PL from
a single V$_{\text{B}}^{-1}$ defect has been reported up to date~\cite{Xu2022-fx, Baber:2023}. \\
\indent 
Several studies have focused on measuring the DC magnetic field sensitivity of V$_{\text{B}}^{-1}$ described by the following
equation~\cite{Gao:2021b,Whitefield, Zhou:2023}:
\begin{equation}
\eta_B=\mathcal{P}_F \times \frac{1}{\gamma_B}\times\frac{\Delta v}{C \sqrt{R}}
\end{equation}
where $\mathcal{P}_F$ is a numerical parameter related to the line shape (for a Lorentzian it corresponds to $0.77 \gamma_B$), and
the electron gyromagnetic ratio equals 28 MHz/mT$^{-1}$. R, $\Delta v$ and C are the emission rate of photons, and ODMR
linewidth and contrast, respectively. The reported sensitivities vary from 100 to 2 $\mu$T/{{Hz}}, reflecting
differences in measured values for ODMR contrast, linewidth, or photon rate. The highest reported
contrast and PL intensity for ensembles of boron vacancies are 46\% and $4.3 \times 10^9$ counts per second, respectively~\cite{Gao:2021b, Zhou:2023}. The ODMR linewidth is influenced primarily by hyperfine coupling to three neighbouring
nitrogen atoms, typically ranging from 100 to 200 MHz~\cite{Zhou:2023}. However, microwave (MW) power broadening can
increase the linewidth up to $\sim $600 MHz~\cite{Gao:2021b}. Nevertheless, the best sensitivity value obtained in these
studies falls short of the performance achieved by leading quantum sensors like the nitrogen-vacancy (NV) centre in
diamond, which typically achieves sensitivities around nanotesla per square root of Hertz (nT/$\sqrt{\text{Hz}}$)
under similar measurement conditions~\cite{Wolf:2023, Balasubramanian2009-hw}.\\
\indent 
\begin{figure*} 
\includegraphics[width=1\textwidth]{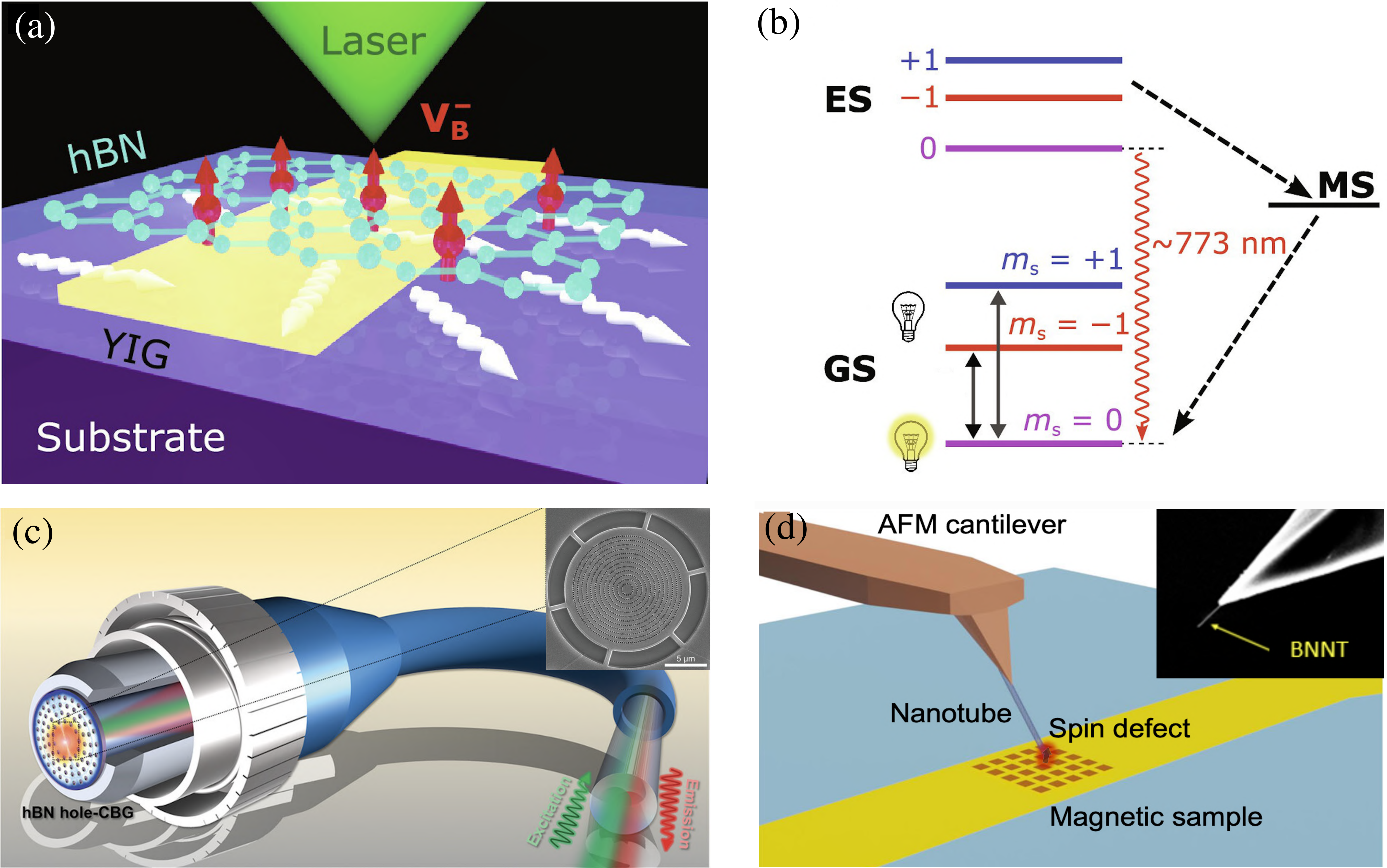}
 \caption{Quantum sensing with spin defects in hBN. (a) A schematic of a hBN nanoflake placed on a gold microwave stripline, which is patterned on a magnetic insulator YIG, designed for quantum sensing measurements. (b) An energy level diagram of a V$_{\text{B}}^{-1}$ defect illustrating the radiative and nonradiative decay processes among the excited state (ES), metastable state (MS), and ground state (GS). The figure (a) and (b) are adapted from Ref.~\cite{Zhou2024-zw} available under a CC-BY-NC 4.0 license. Copyright 2024 Zhou \textit{et al.} (c) Integration of Boron vacancy defects onto a commercial optical fibre for remote sensing. The inset shows the monolithic hBN photonic structure that was transferred on the fibre core. This figure (c) and its subfigure are adapted from Ref.~\cite{Moon:2024} available under a CC-BY-NC-ND 4.0 license. Copyright 2024 Moon \textit{et al.} (d) A single spin defect in Boron Nitride Nanotube attached to an atomic force microscopy tip for high resolution sensing. The inset is an scanning electron microscopy image of the BNNT attached to an atomic force microscopy tip.  This figure (d) and its subfigure are adapted from Ref.~\cite{Gao2023} available under a CC-BY-NC-ND 4.0 license. Copyright 2024 Gao \textit{et al.}\label{Fig:secIII}
}
\end{figure*}
The sensitivity of V$_{\text{B}}^{-1}$ is limited not only by the low brightness, but also by the ODMR linewidth, which is broad compared to other systems. Enhancement of photon emission rate is challenging as the ZPL is symmetry-forbidden, and the position of ZPL is not known~\cite{Qian:2022, Reimers:2020, Ivady:2020, Sortino:2024}. Several works investigated coupling of V$_{\text{B}}^{-1}$ to
photonic structures and achieved a 250-fold enhancement for V$_{\text{B}}^{-1}$ coupled to a plasmonic
nanocavity~\cite{Gao:2021b,Xu2022-fx, Mendelson2021-yv}. Furthermore, strategies to mitigate broadening include engineering of V$_{\text{B}}^{-1}$ in
high-quality hBN crystals, employment of efficient MW antennas, and pulsed excitation techniques.\\
\indent 
Despite the low sensitivity, the hBN host, a two-dimensional crystal, offers distinct advantages, particularly in integration with 2D heterostructures and devices. Additionally, it enables
precise control over thickness down to a monolayer. These features make V$_{\text{B}}^{-1}$ an appealing candidate for applications requiring seamless integration and precise control over the distance between a sample and the sensor.\\
\indent 
Quantum sensing with Van der Waals
heterostructures has been demonstrated by detecting 2D ferromagnets and currents in Graphene~\cite{Huang:2022, Healey:2023}. The relatively large shift in the zero-field splitting of V$_\text{B}^{-1}$ in the 10-295 K temperature range provides
possibilities for simultaneous sensing of magnetic fields and temperature~\cite{Healey:2023}. Moreover, the defect has
been utilized for sensing strain, RF signals, paramagnetic spins in liquids, and spin waves in magnetic materials such as Y$_3$Fe$_5$O$_{12}$ (YIG)~\cite{Zhou2024-zw, Robertson:2023, Gao2023, Patrickson:2024, Rizzato:2023}. Such design on a magnetic material is exemplified in Figure~\ref{Fig:secIII}(a). An energy level diagram of V$_{\text{B}}^{-1}$ spin defect is also demonstrated in Figure~\ref{Fig:secIII}(b), where the optically accessible V$_{\text{B}}^{-1}$ spin states and their dipole interactions with the local magnetic field environment offer an effective method for studying spin wave excitations in a nearby magnetic material. Magnetic field imaging has been demonstrated with a high spatial resolutions of up to
100 nm by integrating hBN onto an array of nanopillars~\cite{Sasaki:2023, Liang:2023}. Additionally, integration of V$_{\text{B}}^{-1}$ onto the tip of a commercial fibre has been achieved by transferring a hBN nanohole Bragg grating cavity onto the centre of the fibre, as shown in Figure~\ref{Fig:secIII}(c)~\cite{Moon:2024}. These advances underscore the adaptability and potential of V$_{\text{B}}^{-1}$ across various sensing domains and integration platforms.\\
\indent 
In parallel, efforts are underway to identify alternative spin defects in hBN. They have led to the discovery of ODMR-active defects in the visible spectral range~\cite{Stern:2022, Mendelson:2021, Exarhos:2019}. These defects can be isolated to
a single level, and have been found to exhibit a broad range of  ZPL emission wavelengths spanning 580 nm to more than 700 nm, presenting
opportunities for interfacing single photons to electron spins. Interestingly, only a fraction of
such emitters in hBN show an ODMR signal and both positive and negative contrast have been observed
in various emitters~\cite{Stern:2022}. The spin system is still under debate, but the spin properties follow spin 1/2 behaviour. As a result, the absence of an intrinsic quantisation axis for these defects has been leveraged for
performing anisotropic magnetic sensing~\cite{Scholten:2024}. They have also been identified in boron nitride
nanotubes and integrated with an atomic force microscopy tip for imaging with high spatial resolution (Figure~\ref{Fig:secIII}(d))~\cite{Gao2023}. \\
\indent 
Spin defects in hBN typically suffer from interactions with the nuclear spin bath of the crystal, leading to reduced spin coherence times. However, a newly-identified spin defect in hBN exhibits surprisingly slow damping of Rabi oscillations. These defects can be isolated to a single level and feature a ground state triplet with a zero-field splitting of 1.96 GHz at room temperature. Although the initial spin coherence time recorded for these defects was around 100 ns, it was increased to 1 $\mu$s using 10 focusing pulses in a multipulse dynamical decoupling scheme~\cite{Stern:2024}. Achieving such extended coherence at room temperature highlights the potential of hBN for quantum hardware development, particularly for quantum communications and quantum networks.\\
\indent 
Research on both V$_{\text{B}}^{-1}$ and visible defects in hBN continues, with distinct emphasis on improving
sensitivity and understanding fundamental properties, respectively. Efforts to enhance V$_{\text{B}}^{-1}$
sensitivity involve leveraging advanced measurement techniques and integration with nanophotonic architectures~\cite{Sortino:2024, Rizzato:2023}. Meanwhile, fundamental questions surrounding the electronic level
structure and ODMR mechanism of visible hBN defects remain unanswered. Moreover, the lack of
capability in on-demand creation of visible defects hinders their application in quantum sensing. However, once these challenges are overcome, these defects hold significant promise for quantum sensing applications, presenting numerous potential opportunities for advancements in the field.

\subsection{Extended Quantum Theory Test Experiments}
\begin{figure}[h!]
    \centering
    \includegraphics[width=\linewidth]{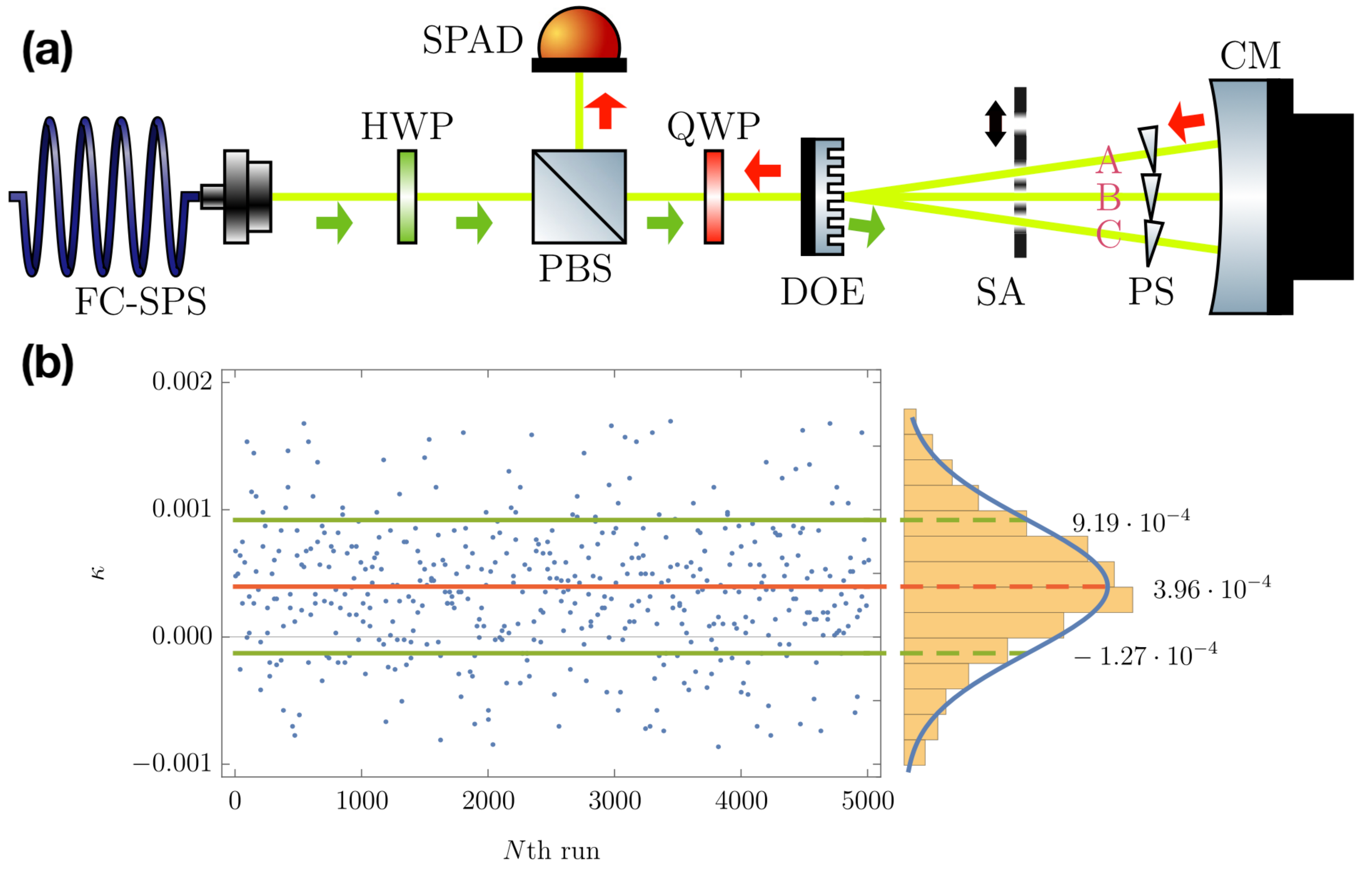}
    \caption{{{(a) Experimental setup for higher-order interference testing. A single-photon source's polarization is rotated and split into three paths using a beam splitter, then recombined for interference measurement. Optional phase shifters adjust path interference, and shutter arrays control individual path openings. (b) Results of interference testing, showing averaged measurements of the parameter $\kappa$ across 5000 runs, with statistical variation presented as a histogram. Here, $\kappa$ describes the deviation from Born's rule, specifically quantifying the third-order interference compared to the expected second-order interference in the three-path interferometer experiment. This figure is adapted from Ref.~\cite{TV-PRB2021} available under a CC-BY 4.0 license. Copyright 2021 Vogl \textit{et al.}} }}\label{Fig:test}
\end{figure}

This section outlines how the properties of hBN single-photon emitters have been applied to investigate certain concepts in quantum theory. \\
\indent 
Recently, the potential for exploring quantum phenomena using hBN as a single-photon emitter~\cite{Vogl:2019-QC} was examined in a proposal in Ref.~\cite{TV-PRB2021}. In particular, the study explored how a defect center in hBN, as a quantum light source, could be used to experimentally investigate higher-order interference, specifically focusing on potential deviations from Born’s rule {{(see Figure~\ref{Fig:test} for the experimental setup)}}.

\indent 
Some of the earliest works in the quantum theory test literature, predating the quantum light source perspective that now shapes the language of applications, already noted that deviations from established postulates within quantum theory could yield observable consequences in experimental settings~\cite{Peres1979, Peres1979, Kaiser-PRA1984, Sorkin1994}. For example, deviations from Born's rule can result in higher-order interference~\cite{Sorkin1994}, while in hypercomplex Hilbert spaces, phase commutativity may be disrupted~\cite{Peres1979}. More recently, proposed experimental tests of these extended quantum theories have progressively constrained the potential for deviation by testing hypercomplex quantum mechanics with relativistic photons~\cite{Procopio2017} and higher-order interference using multipath interferometers~\cite{Urbasi2010,Soellner_2011,Kauten2017,Pleinert2020,Pleinert2021}. However, common challenges persist, particularly in accurately characterizing complex measurement apparatuses, which can introduce noise and systematic errors that may lead to apparent deviations. Additionally, quantum fluctuations in the light source and the nonlinear response of photodetectors impose fundamental limits on measurement accuracy~\cite{Kauten2017}, as uncertainties in apparatus characterization inevitably translate into uncertainties in observed outcomes. \\
\indent 
The use of a single-photon source offers a different approach compared to prior tests by addressing common experimental limitations, such as shot noise and detector nonlinearity. By spacing consecutive single photons in time, this method helps minimize intensity fluctuations and ensures a more linear detector response, improving upon traditional detection methods like single-photon avalanche diodes. As a result, this approach allowed for the establishment of a limit on higher-order interference, which was measured at $\kappa=3.96(523)\times10^{-4}$ relative to the expected second-order interference. It is worth noting, that in general the room temperature emission capabilities are not necessary for such tests, so semiconductor quantum dots could even achieve a better sensitivity than the hBN source in the previous experiment~\cite{TV-PRB2021}.

\subsection{Quantum Key Distribution} 
\begin{figure*}[ht]
\includegraphics[width=.54\textwidth]{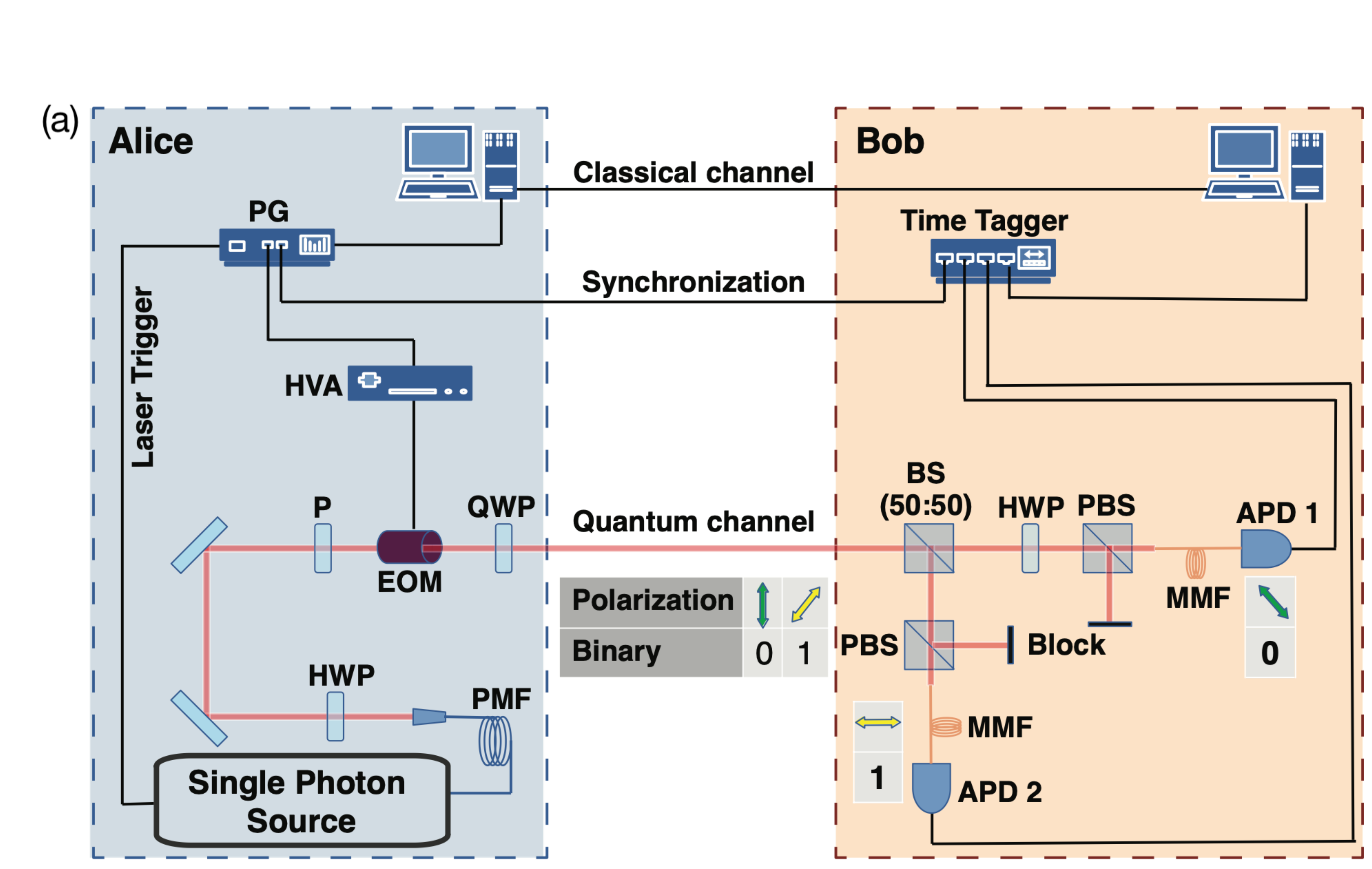} 
\includegraphics[width=.44\textwidth]{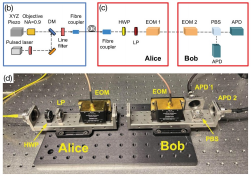} 
    \caption{Experimental setups for the first proof-of-concept QKD demonstrations employing single photon emission in hBN. (a) The experimental setup of B92-based free-space QKD system performed in Ref.~\cite{Samaner2022}, and this figure is adapted from Ref.~\cite{Samaner2022} available under a CC-BY 4.0 license. Copyright 2024 Samaner \textit{et al.} Alice side is the pulse generator that drives the photon source at 1 MHz, synchronizes with the time-tagging module, and controls the EOM to modulate polarization to V (0) or +45 (1), while Bob side is a beam splitter that directs photons to APD1 (-45, 0) or APD2 (H, 1), with detection times recorded for reconciliation. (b) A schematic of the single-photon source, (c) a schematic of the QKD setup (EOM: electro-optic modulator; LP: linear polarizer; APD: avalanche photodiode; PBS: polarizing beam splitter; HWP: half-wave plate.), and (d) the optical components of the transmitter (Alice) and receiver (Bob) summarize the implementation based on the BB84 QKD protocol performed in Ref.~\cite{Juboori2023}. The Figure (b), (c) and (d) are adapted from Ref.~\cite{Juboori2023} available under a CC-BY-NC 4.0 license. Copyright 2023 Juboori \textit{et al.}  
 }
\label{Fig:secQKD}
\end{figure*}
\begin{table*}[htbp]
    \centering
    \begin{tabular*}{\textwidth}{@{\extracolsep{\fill}}ccllrllrc}
    Source&Ref.&Protocol&Encoding&$\lambda$(nm)&Photon per pulse&QBER(\%)&SKR/Freq.(Bitsx$10^{-6}$)&Bound(dB)\\\hline
    \multirow{3}{*}{hBN}&\cite{Zeng2022}& BB84&Passive&728&-&-&-&26\\
    &\cite{Samaner2022}& B92&Pol.&671&0.011-0.023&8.9-11.3&(238-414)x$10^{-6}$&22\\
    &\cite{Juboori2023}& BB84&Pol.&645&0.012&3.0-8.0&(504-900)x$10^{-6}$&-\\\hline
    \multirow{3}{*}{QDs}&\cite{Waks2002}&BB84&Pol.&877&0.007&2.5&329x$10^{-6}$&28\\
    &\cite{Rau2014}&BB84&Pol.&910&0.0034&6.0-9.0&33.6x$10^{-6}$&-\\
    &\cite{Takemoto2015}& BB84&Pha.&1580&0.05&2.3&70x$10^{-6}$&23\\\hline
    \multirow{2}{*}{NV Center}&\cite{Beveratos2002}&BB84&Pol.&637-750&0.014&4.6&1792x$10^{-6}$&20\\
    &\cite{Leifgen2014}&BB84&Pol.&600-800&0.029&3.0&2600x$10^{-6}$&16\\\hline
    TMDCs&\cite{Gao2023}&BB84&Passive&807&0.013&0.8&-&23\\\hline
    Molecule&\cite{Murtaza2023}& BB84&Passive&780-830&0.04-0.08&2.0-3.9&-&23-27\\\hline
    \end{tabular*}
    \caption{Comparison of QKD demonstrations using color centers in  hBN and other single photon sources. (SKR: Secure Key Rate).}
    \label{tab:qkdcomparison}
\end{table*}
Quantum Key Distribution (QKD) is a method of secure communication that uses principles of quantum mechanics to enable two parties creating a shared random secret key to be used for encrypting and decrypting messages. QKD ensures security by detecting any attempt at eavesdropping on the communication channel, as any attempt to measure the qubits being transmitted would create a detectable trace in the quantum channel. The first proposed QKD protocol is known as BB84~\cite{Bennett2014} and, to date, various types of protocols have been put forward including prepare and measure (PM-QKD) \cite{Bennett2014,Bennett1992}, entanglement-based (EB-QKD) \cite{Ekert1991,Bennett1992BBM} and measurement device independent (MDI-QKD) \cite{Lo2012,Lucamarini2018}. Here, we focus on demonstrations of PM-QKD, which relies on polarization encoding of single photons as flying qubits. Employing weak coherent pulses as the light source offers a practical but limited (typically 0.05 photons per pulse) platform for implementing PM-QKD. Although the decoy-state method~\cite{Lo2005} beats the secure key rate that can be achieved using weak coherent pulses, an efficient single-photon source outperforms weak coherent pulses QKD in the same average photon number \cite{Waks2002-2}. Therefore, the utilization of single-photon sources in QKD systems is crucial and keeps gaining momentum, driven by enhancements in performance parameters such as purity and brightness. 
\\
\indent 
So far, several single-photon sources have been used in proof-of-concept QKD demonstrations such as quantum dots (QDs) \cite{Waks2002,Intallura2007,Takemoto2010,Takemoto2015,Heindel2012,Rau2014,Zahidy2024}, color centers in diamond \cite{Beveratos2002,Alleaume2004,Leifgen2014}, transition metal dichalcogenides (TMDCs) \cite{Gao2023} and single dibenzoterrylene molecules (DBT) \cite{Murtaza2023}. QDs and TMDCs are good candidates for QKD applications, but their requirement of cryogenic temperature limits their potential in practical applications. On the other hand, diamond color centers operate at room temperature but compared to the other single-photon sources they have lower brightness potential and higher lifetime, limiting high key rate generation in QKD applications.\\
\indent 
Defects in hBN stand out among other solid-state single photon sources with high brightness and purity at room temperature. Following initial investigations on their potential in QKD \cite{Zeng2022}, the first proof-of-concept QKD demonstration with bright single photon emission in visible band from defects in hBN has been reported \cite{Samaner2022}. In their work, Samaner and colleagues employed the B92 QKD protocol using two non-orthogonal polarization states for active encoding at 1 MHz, and they achieved a sifted key rate of 238 bits/s with a quantum bit error rate (QBER) of 8.95\%. The setup used in this experiment is presented in Figure~\ref{Fig:secQKD}(a). Implementation of BB84 QKD in free-space with defects in hBN has also been demonstrated \cite{Juboori2023}, with 450 bits/s as the key rate and 5\% QBER under 500 kHz operation, for which the experimental setup is shown in Figure~\ref{Fig:secQKD}(b), (c), (d). The source brightness is one of the most important parameters to reach high key rates, which can be increased through Purcell enhancement by coupling emitters to plasmonic antennas or dielectric cavity-based structures \cite{Tran2017,Haubler2021, Vogl:2019-QC}, and can help beat the decoy-state protocol \cite{Zhang2024}. In addition, controlling the lifetime may extend the power of temporal filtering, which is an important tool for reducing QBER and maximizing the communication distance, especially in noisy channels \cite{Ko2018,Kupko2020,Samaner2022}. Another important source parameter for efficient QKD is the wavelength of the single photon emission that determines the transmission efficiency of photons in different mediums. Due to its wide bandgap, hBN hosts various defects emitting single photons from UV to NIR \cite{Cholsuk2024a} that can be chosen for best transmission in different quantum channels. Recently, B-centers in hBN emitting at 436 nm have been used for BB84 QKD in water with similar rates and QBER values as obtained in air \cite{Scognamiglio2024}. \\
\indent 
To underscore the significance of these findings, table~\ref{tab:qkdcomparison} lists important parameters for single photon based QKD. Experiments indicated by passive encoding represent the results obtained by incorporating the necessary experimental parameters into the GLLP equation \cite{GLLP2004}. The listed QBER values are clearly seen to be very different from each other. It should be noted that QBER values are highly dependent on QKD experimental setups, and only a small amount of these errors are related to the photon source. Therefore, it is difficult to compare single photon sources by taking QBER values into account. The maximum tolerable QBER values for the BB84 and B92 protocols have been reported as 11.2 \cite{Lutkenhaus2000} and 15\% \cite{Bunandar2020} (B92 security limit depends on the angle between non-orthogonal states), respectively. Although single photon emitters based on defects in diamond or single molecules are brighter compared to others, the broad wavelength range increases the error rate as it passes through wavelength-dependent components such as the electro-optic modulator. Finally, it can be said that experiments involving
hBN are still in the early stages; nevertheless, the secret key rate per pulse and bound values are comparable or even better than those of others. Therefore, future experiments are expected to be more promising.

\subsection{Quantum Computing}

\begin{figure*} 
\includegraphics[width=.465\textwidth]{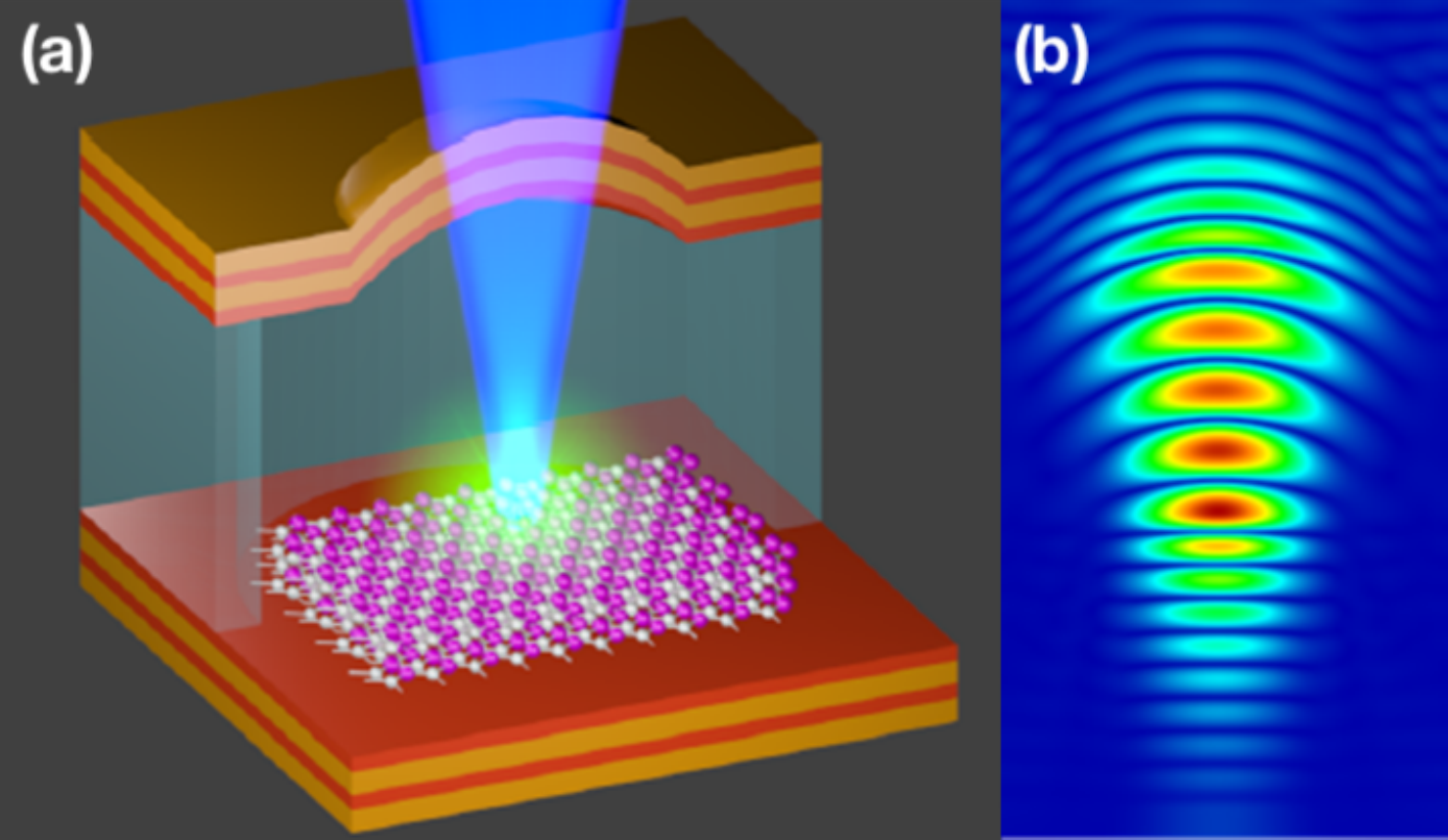} \hspace{+.1cm}
\includegraphics[width=.48\textwidth]{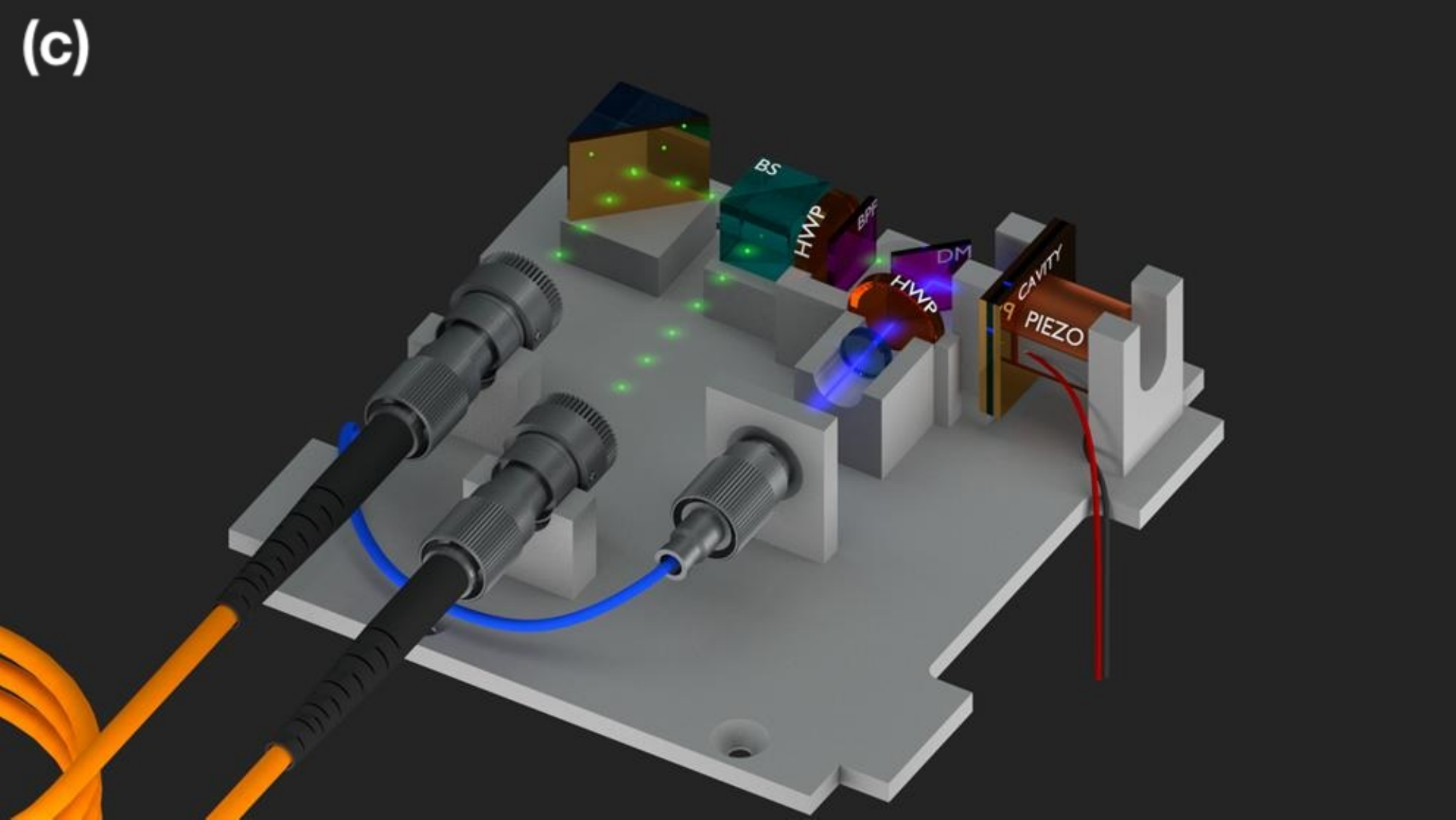} \\
\includegraphics[width=.49\textwidth]{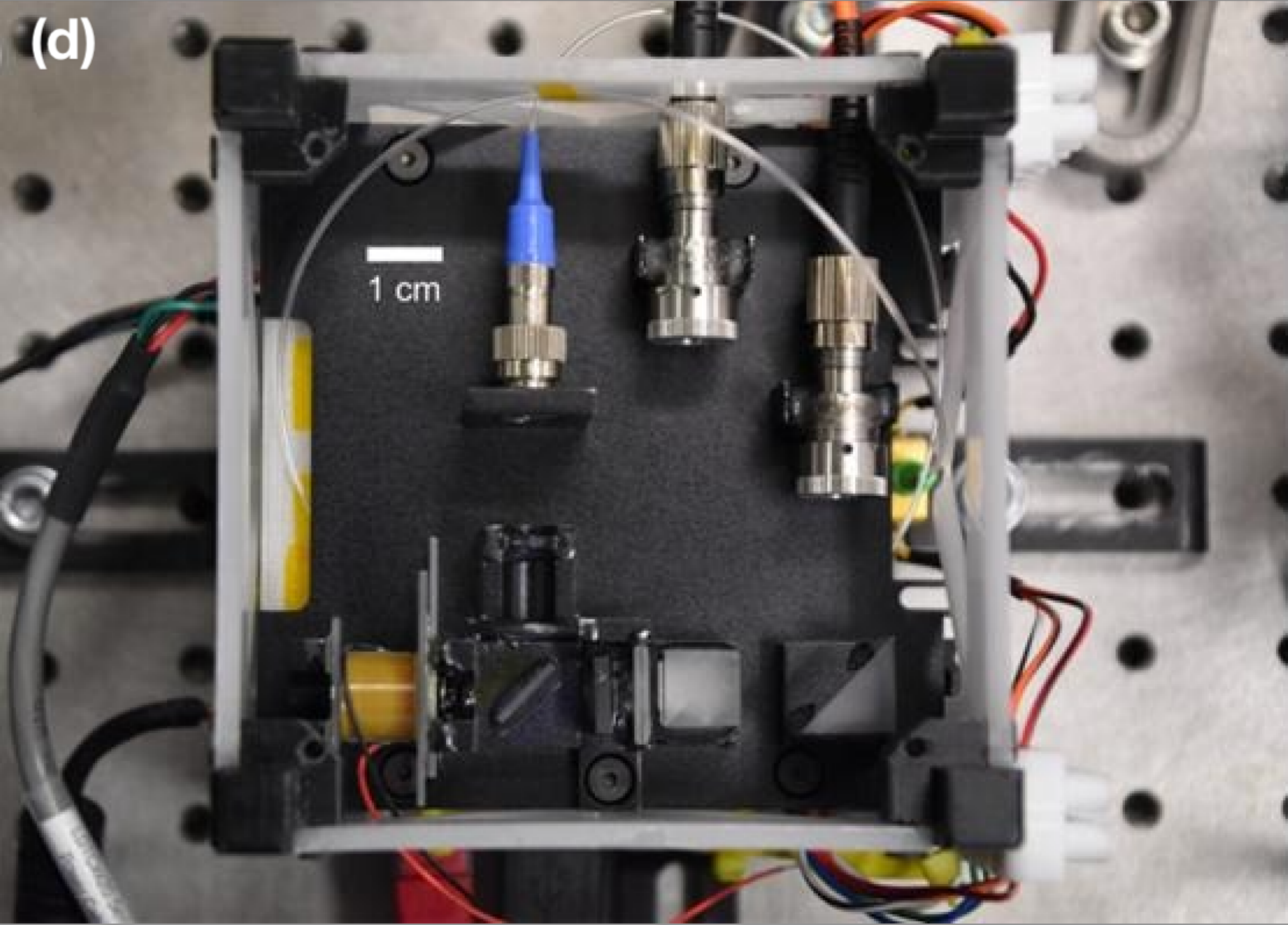} \hspace{+.1cm}
\includegraphics[width=.37\textwidth]{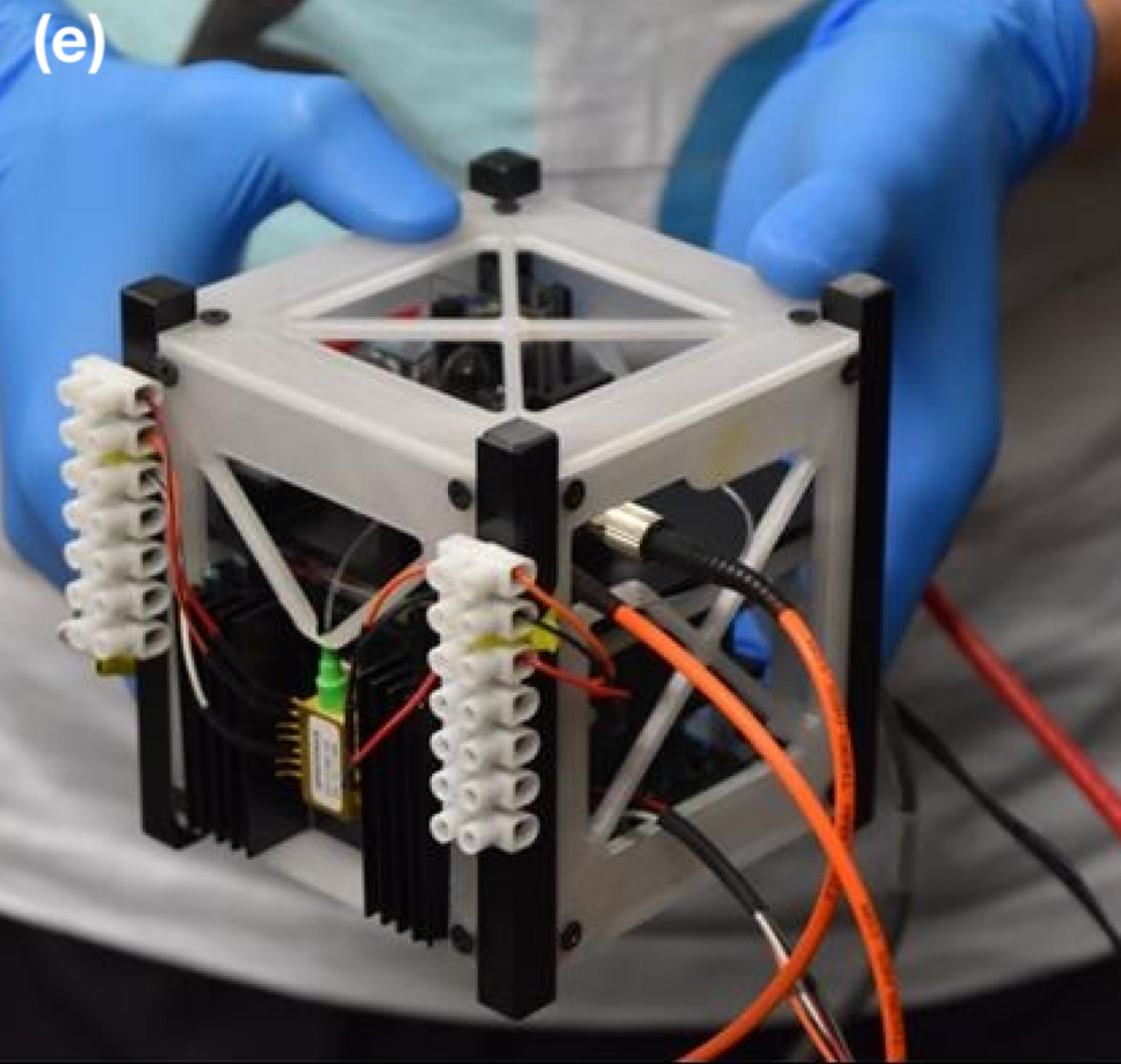} 
	\caption{(a) The microcavity consists of a hemispherical and flat mirror. The quantum emitter emits confocally with the excitation laser, and a PDMS spacer controls the cavity length, with the center etched to avoid interference with the emitter. (b) Result of the finite-difference time-domain simulations performed in Ref.~\cite{Vogl:2019-QC}. As the cavity length approaches resonance, the intensity increases, and (b) represents at resonance when the intensity reaches its maximum. (c) Design of the optics platform. A polarization-maintaining fiber directs the excitation laser into the cavity, focusing it onto the defect. The single photons pass through a dichroic mirror, are filtered, then split by a beam splitter and coupled into multimode fibers. (d) Top, and (e) Side view photographs of the device taken from Ref.~\cite{Vogl:2019-QC}. The frame is modeled after a 1U CubeSat, a small satellite platform in the pico-class category. Figures are adapted with permission from Ref.~\cite{Vogl:2019-QC}. Copyright 2019 American Chemical Society.
}
\label{Fig:secQC}
\end{figure*}

\begin{figure*} 
    \centering
    \includegraphics[width=.9\linewidth]{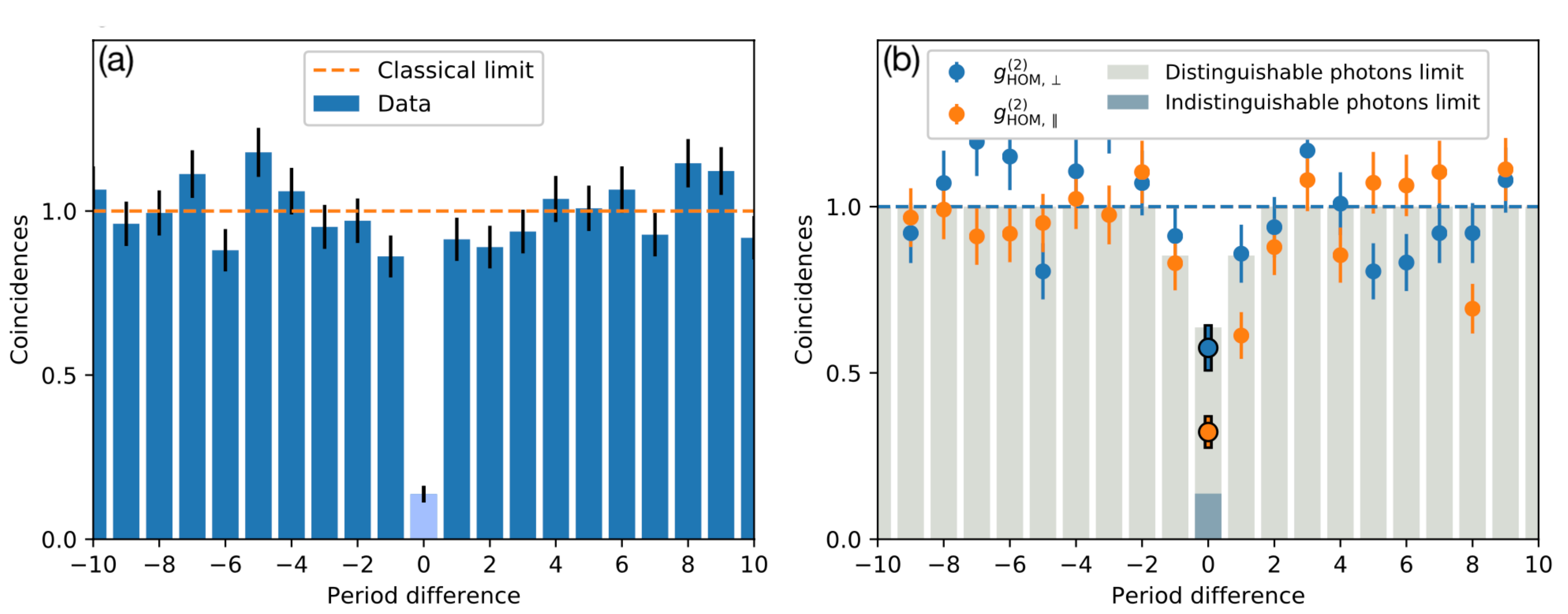}
    \caption{(a) The second-order correlation function \( g^{(2)}(0) \), which indicates the presence of single-photon emission. The low value of \( g^{(2)}(0) = 0.14 \) suggests the photons exhibit strong antibunching, confirming that the source emits photons one at a time rather than in bunches. (b) Results from a Hong-Ou-Mandel interference experiment in Ref.~\cite{Fournier:2023}, which measures two-photon coincidences for both parallel and orthogonal polarizations. The difference in coincidence rates for the two configurations helps determine the degree of photon indistinguishability. Lower coincidence rates in the parallel case (0.32) compared to the orthogonal case (0.58) suggest partial indistinguishability of the photons. Figures are reprinted with permission from Ref.~\cite{Fournier:2023}. Copyright 2023 by the American Physical Society. 
}\label{Fig:secQC2}
\end{figure*}
Quantum computing, particularly through superconducting, trapped-ion, and photonic devices, plays a crucial role in advancing quantum metrology and quantum information tasks. These platforms enable precise multi-parameter estimation through collective measurements, leveraging quantum entanglement and coherence to achieve precision unattainable by classical resources. This makes them especially valuable in scenarios where resources are limited, as quantum-enhanced metrology can maximize the information gained from minimal data~\cite{Parniak:2018, Hou2018-jw, Wu:2020, Yuan2020-so, Jurcevic_2021, Pogorelov:2021, Conlon2023-ij}.\\
\indent 
Different platforms exhibit distinct strengths and limitations, which underscores the diverse capabilities and ongoing development needs across various quantum computing technologies: Superconducting devices, such as IBM's F-IBM QS1~\cite{Jurcevic_2021}, relying on error mitigation techniques to achieve high precision particularly in multi-qubit operations, make them suitable for complex quantum tasks despite their susceptibility to gate and readout errors. In contrast, trapped-ion processors like AQTION~\cite{Pogorelov:2021} demonstrate exceptional precision in single qubit operations without needing error mitigation, though they face challenges in scaling up to multi-copy measurements. The third front-runner among the quantum computing platforms are photons with the Jiuzhang device~\cite{doi:10.1126/science.abe8770}. Photonic quantum computers are limited by multi-qubit operations as photons do not directly interact. This can be circumvented by measurement-based quantum computing~\cite{PhysRevLett.86.5188}, where a large entangled cluster state is used and then only single qubit operations are required. This makes quantum computing interesting for single photons emitted from hBN, even though no entangled cluster states have been generated from hBN so far.\\
\indent 
The pioneering work utilizing single qubit operations with a single photon emitting color center in hBN was first introduced in Ref.~\cite{Conlon2023-ij} by the JenQuant device, exemplifying the integration of quantum computing with advanced photonic technology. JenQuant makes use of the particular hBN defects coupled with a microcavity with a ZPL at 565.85 nm and a Lorentzian linewidth of 5.76 nm. These defects are created using oxygen plasma etching followed by rapid thermal annealing under an argon atmosphere, which forms defects primarily around 560 nm~\cite{Vogl:2019-QC}. These quantum emitters hosted by multilayer hBN flakes are then integrated into a tunable optical microcavity (Figure~\ref{Fig:secQC}), which enhances their performance by improving emission directionality and noise suppression through the Purcell effect. This showcases significant advancements in photonic quantum computing:
i) Photon collection efficiency; enhanced emission rate and directionality improves photon collection efficiency, which is vital for readout processes in quantum computing. ii) Noise reduction; enhanced emission also reduces the probability of multi-photon events, minimizing errors in quantum computations. In addition, hBN's ability to tune the emission spectrum allows for matching with specific wavelengths required by different quantum computing and communication components, such as single-photon detectors and photonic circuits. Tunable emitters can be integrated seamlessly with photonic chips and waveguides, facilitating the development of scalable quantum photonic circuits. Further, the stable emission properties of hBN emitters support the implementation of error mitigation techniques, improving the fidelity of quantum operations.  
All of these make the hBN highly compatible with various components of quantum computing systems mentioned above, supporting the development of reliable, efficient, and scalable quantum technologies.\\
\indent
In JenQuant, logical qubits are encoded in photon polarization with $|H/V\rangle$ as the computational basis states $|0/1\rangle$, achieving high polarization extinction ratios ($>10^5$). This setup, which implements single-copy positive operator-valued measurements (POVMs) through motorized polarization optics, ensures precise control over quantum states.  The experiments conducted~\cite{Conlon2023-ij} show that when compared to trap-ion processor AQTION~\cite{Pogorelov:2021}, both JenQuant and AQTION excel in reaching the theoretical single-copy measurement limits without error mitigation. JenQuant's strength lies in its inherent stability and efficiency for single qubit operations, while AQTION showcases high precision and the ability to perform effectively in environments with significant decoherence, though it does not reach the two-copy measurement limits as JenQuant does.
Compared to IBM Quantum's F-IBM QS1 device~\cite{Jurcevic_2021}, which uses superconducting qubits, was tested for simultaneous qubit rotation estimation (IBM Quantum Falcon processor), JenQuant photonic processor achieves high-precision single qubit operations naturally due to its inherent stability without the need for error mitigation as mentioned above, whereas the IBM F-IBM QS1 device requires significant error mitigation techniques to reach similar levels of precision~\cite{Conlon2023-ij}.
This comparison highlights the strengths of each platform: JenQuant excels in stability and single-qubit functionality, while IBM Quantum devices achieve high precision in multi-qubit operations with error correction. While JenQuant currently achieves only single qubit gate operations, the use of entangling gates, like controlled NOT gates, is promising for enabling multi-qubit operations. Unlike single qubit gates, which do not involve interactions between qubits, entangling gates facilitate these necessary interactions. 
Such gates can be achieved with resonator linewidths narrower than 124 MHz, resulting in a Hong-Ou-Mandel (HOM) contrast $>$90\%, thereby improving photon indistinguishability~\cite{Vogl:2019-QC}. So far, HOM inteference from quantum emitters in hBN has been demonstrated at low temperature~\cite{Fournier:2023} (see Figure~\ref{Fig:secQC2}). With \( g^{(2)}(0) = 0.14 \) and HOM dips of 0.32 and 0.58 for parallel and orthogonal polarizations, respectively, have been obtained from a single hBN emitter~\cite{Fournier:2023}.\\
\indent 
Combination of the strengths of quantum computing and hBN-based photonic technology underscores the significant practical advantages in applications such as biomedical imaging, quantum communications, and entanglement distillation. The demonstrated stability and precision in hBN-based systems, along with the potential for multi-qubit operations, illustrate a promising path forward for near-future quantum technologies.

\subsection{Quantum Random Number Generation}
\begin{figure*}
    \centering
    \includegraphics[width=0.41\linewidth]{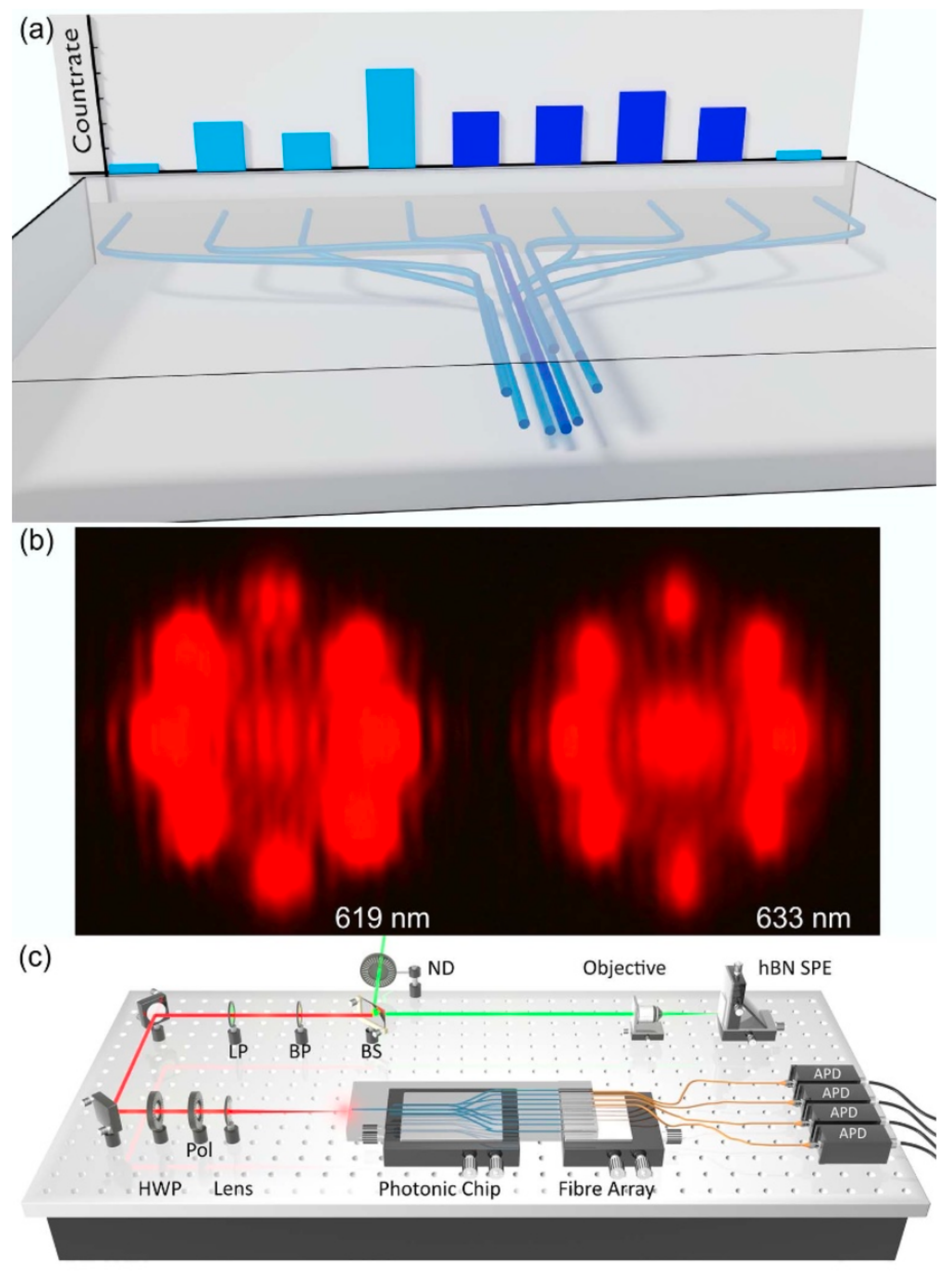}
    \includegraphics[width=0.49\linewidth]{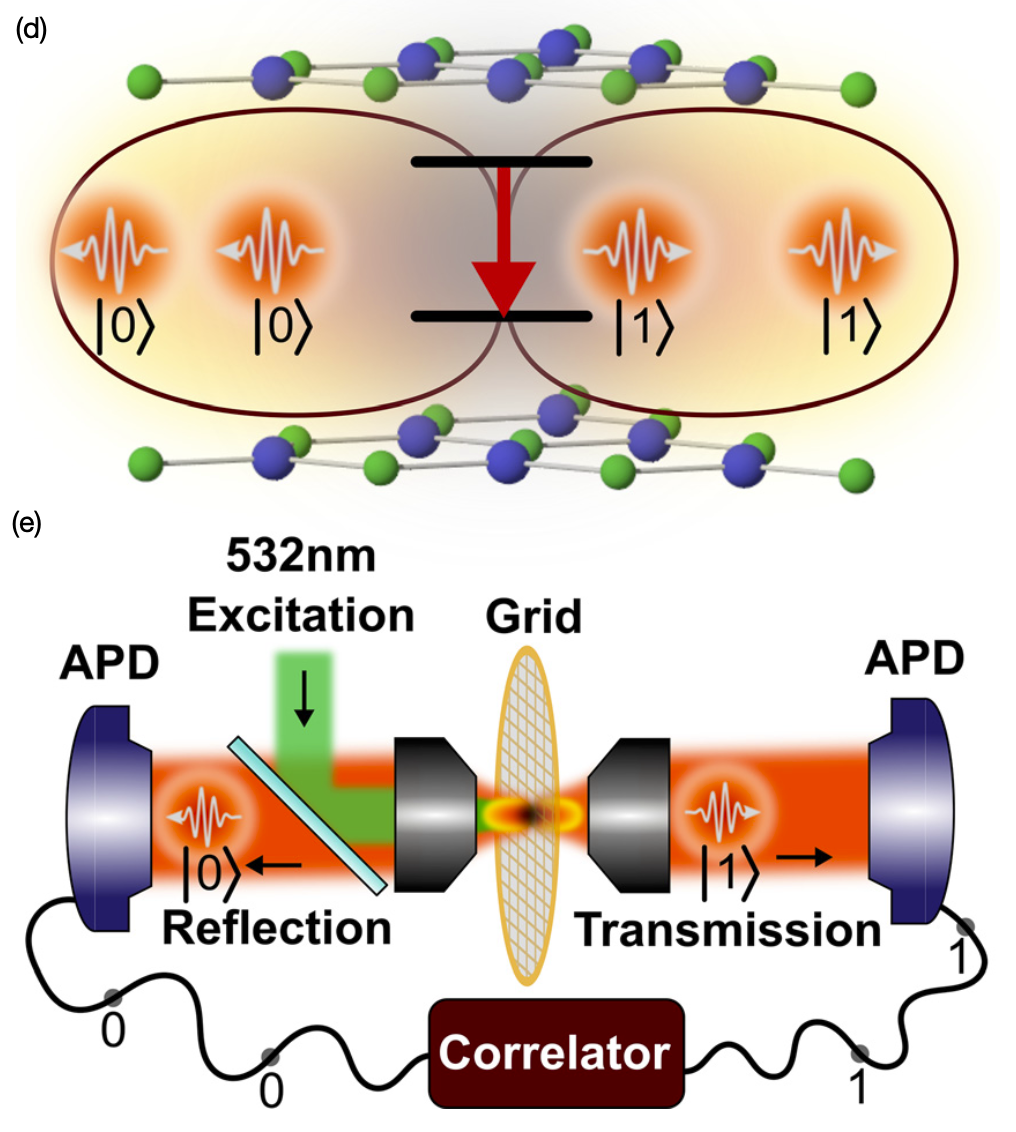}
    \caption{{{(a) A single input waveguide channels light into eight radial waveguides that spread out for output; photon counts across four selected waveguides show a standard deviation of 11.6$\%$. (b) Chip splitting efficiency shown at 619 and 633 nm wavelengths. (c) Chip setup overview: a 532 nm laser excites the hBN emitter, and photons are collected through APDs after adjusting polarization with waveplates and filters. Figures (a), (b) and (c) are reprinted with permission from Ref.~\cite{White_2021}. Overview of the experimental setup: (d) Quantum randomness is generated by encoding photon directions as binary states (0 and 1), exploiting the symmetry in the dipole emission pattern. (e) The setup features a defect center in hBN excited at 532 nm, with emitted photons detected in reflection and transmission channels using APDs. Figures (d) and (e) are adapted from Ref.~\cite{Hoese2022} available under a CC-BY 4.0 license. Copyright 2022 Hoese \textit{et al.}}
}}\label{Fig:QNRG1}
\end{figure*}

Another crucial concept and resource for applications in cryptography, secure communications, and advanced computing is quantum random number generation (QRNG), which makes use of the fundamental principles of quantum mechanics to produce truly random numbers~\cite{OBrien2009-oq}. Unlike classical random number generators~\cite{knuth97}, which rely on deterministic processes and can be potentially predicted~\cite{Herrero:2017}, QRNG exploits the intrinsic randomness of quantum phenomena to ensure unpredictability and security. \\
\indent 
Recent advancements in the use of hBN as a material platform for QRNG have shown promising results~\cite{White_2021, Hoese2022} with innovative approaches to harnessing single-photon emissions from hBN. Ref.~\cite{White_2021} 
demonstrates a successful implementation of QRNG using a hBN single-photon emitter integrated with an on-chip photonic circuit {{(see Figure~\ref{Fig:QNRG1}(a), (b), and (c) for schematic of the chip)}}: The hBN emitter was prepared by dispersing multilayer flakes on a silicon substrate, followed by annealing. The system was excited using a 532 nm laser, emitting photons through a photonic chip designed in a wagon-wheel structure with multiple waveguides.
The emitted photons were filtered, focused, and coupled to the chip, which then distributed the photons across multiple output channels. Detection was achieved using avalanche photodiodes with high efficiency and low dark counts.
As for the single-photon purity, it was verified using second-order correlation measurements, confirming a low g$^{(2)}$(0) value, indicative of true single-photon emission.
The system's efficiency was demonstrated with a photon count rate of around 1 MHz before the chip and a throughput of approximately 350 kHz after the chip. This setup not only ensures high purity and efficiency of photon generation but also offers a robust platform for scalable quantum photonic technologies, as more bits per photon can be achieved by increasing the number of waveguides
beyond the limitations of traditional bulk-optical arrangements, thereby increasing the number of bits that can be encoded per photon. The idea behind is to exploit the phenomenon where single photons, emitted by the hBN defects, are randomly distributed across multiple output channels in a waveguide structure. 
\\
\indent 
When a single photon is emitted, it enters one of the designated output channels in a fundamentally random manner, without any predictable pattern. This process harnesses the inherent unpredictability of quantum mechanics, as the exact path the photon takes is not determined until it is measured. The random number generation scheme records the output channel of each detected photon, using the occurrence of photons in different channels to generate a sequence of random binary numbers. This randomness is further verified through the National Institute of Standards and Technology (NIST) randomness test suite~\cite{Bassham:2010}, ensuring the unpredictability and statistical quality of the generated numbers.\\
\indent 
On the other hand, Ref.~\cite{Hoese2022} handles the implementation of QRNG with the use of a symmetric dipole emission pattern from defect centers in hBN to generate randomness {{(see Figure~\ref{Fig:QNRG1}(d) and (e) for the experimental overview)}}: QRNG setup utilizes special types of emitters within hBN that exhibit coherent interactions under ambient conditions, demonstrated by the distinct ZPL at 667 nm. These specific defect centers are located in multi-layer hBN and are characterized by mechanical decoupling from low-energy phonon modes, evidenced by the energy splitting between the ZPL and the phonon sideband (PSB). The emitters exhibit a high visibility of the emission dipole, indicating pure dipolar emission with negligible higher-order contributions. These characteristics enable the use of these hBN defects in generating high-quality random numbers, as they ensure the statistical independence and intrinsic randomness necessary for QRNG applications.
\\
\indent 
This experiment confirms that single photons are emitted with equal probability in forward and backward directions, corresponding to the reflection and transmission channels, respectively. Photon detection channels are nearly balanced, and second-order correlation measurements validate the single-photon nature of the emitters, with the setup capable of generating high-quality random numbers, as confirmed by the NIST randomness test suite~\cite{Bassham:2010}. As for the alignment of the hBN flakes and defect centers, it was achieved using atomic force microscopy to manipulate the flakes, ensuring that the mechanical isolation of the defect centers remains intact, maintaining the quality of photon emission. \\
\indent 
These studies together demonstrate the progress made in utilizing hBN for QRNG, illustrating feasibility of solid-state QRNG systems, and the potential for developing high-quality, compact QRNG chips that can be integrated into laptops, smartphones, and other consumer electronics. This technology could greatly enhance the security features of everyday devices by employing the unique properties of hBN-based quantum emitters, offering robust, hardware-based cryptographic solutions. Interesting directions for future works could include addressing challenges such as spectral diffusion and phonon interactions, which influence emission properties significantly, and further investigation into optimizing these aspects through nanomanipulation and the integration of photonic devices, which could significantly boost the performance of QRNGs and facilitate their incorporation into broader quantum technology systems.
Another avenue for future research could involve exploring hBN defects outside the currently studied range of ZPL, along with increasing the number of waveguides which could enable a single photon to generate a larger number of random bits, making it a highly promising solution for integrated quantum photonics and quantum information processing.
\subsection{Quantum Memory}
Numerous applications of hBN single photon emitters have been introduced in the last years, yet there are a few tools that are used to obtain state-of-the-art results in quantum memories: 
Here, the potential implementations of defects in hBN for quantum memories are discussed in comparison with other physical systems. After outlining the fundamentals of quantum memory, the discussion proceeds with two commonly employed quantum memory mechanisms, which are Raman-based and nuclear-spin solid-state quantum memory models. Finally, state-of-the-art research and outlook are provided.\\
\indent 
A crucial element in photonic quantum technologies, a quantum memory, denotes a system with the ability to reliably store quantum states, preserving them for a certain period of time, and transmitting them back to successive communication nodes. This potentially gives rise to possible large-scale quantum technologies, not only for quantum communication \cite{10.1038/s41534-021-00460-9,10.1038/srep20463}, but also for quantum computing \cite{10.1049/iet-qtc.2020.0002}.\\
\indent 
Quantum memory efforts based on defects in hBN and NV centers in diamond have been studied in succession under different constructions. Thus, the aim is here to provide a comprehensive picture, examining potential of hBN-based quantum memory alongside its comparison with diamond-based counterpart.
\stoptocentries
\subsubsection{Raman-based solid-state quantum memory protocol}
\begin{figure}[ht!]
    \centering
    \includegraphics[width=\linewidth]{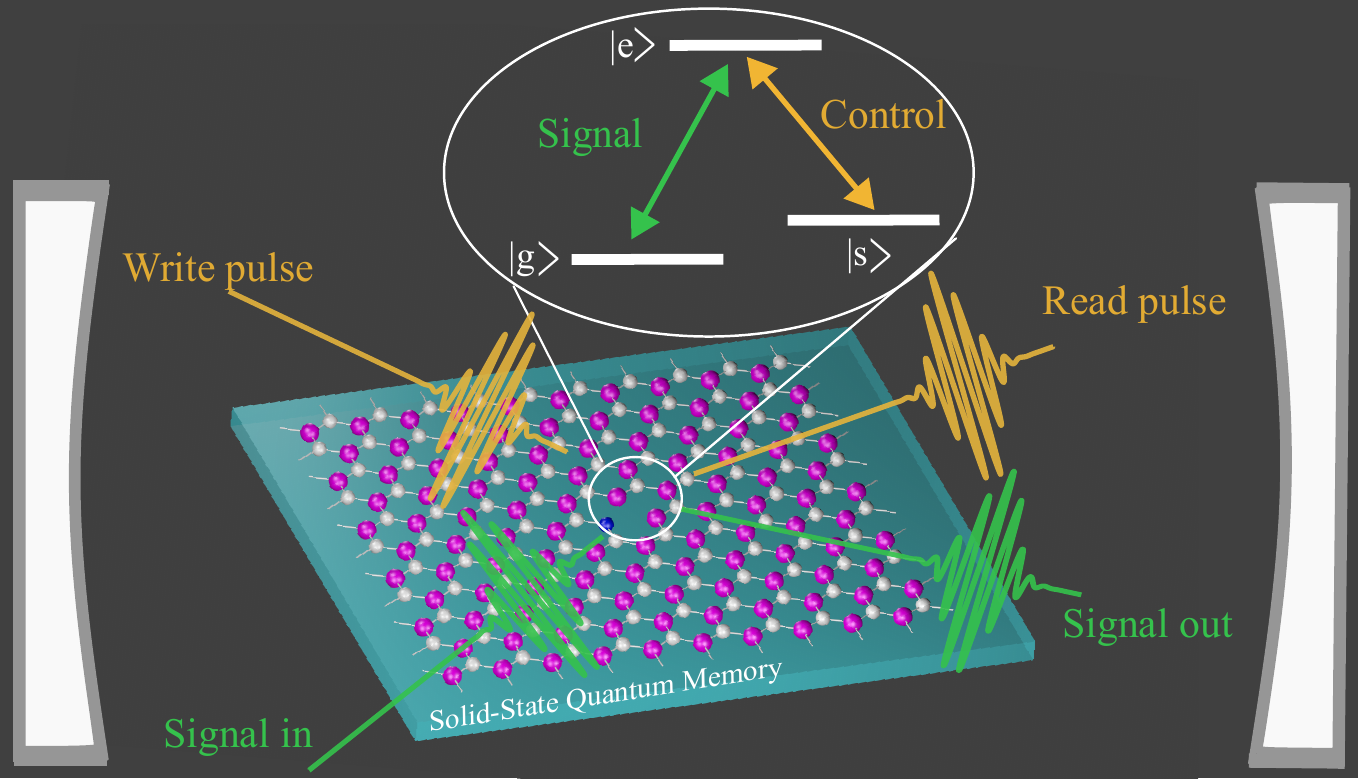}
    \caption{$\Lambda$ structure in hBN potentially used in Raman-based quantum memory protocol where it consists of ground $\ket{g}$, excited $\ket{e}$, and meta-stable $\ket{s}$ states. The transition is operated by two pulses; signal pulse for coupling with the incoming photons and control pulse for write/read process. The sample is embedded in an optical cavity.}
    \label{fig:lambda_structure}
\end{figure}

\indent 
At the theoretical level, the initial implementation of the Raman quantum memory scheme was first conducted in~\cite{10.1103/PhysRevA.89.040301} using negatively charged NV centers in diamond, after which it was subsequently applied within the hBN framework. The first theoretical investigations, in which the ability of defects hosted by hBN, are introduced very recently in \cite{arXiv:2306.07855,10.1002/adom.202302760}. These studies have, in parallel, made use of a cavity-enhanced off-resonant Raman quantum memory protocol. 
In this protocol, a solid-state system should, in principle, feature the $\Lambda$ electronic structures, consisting of a ground state $\ket{g}$, an excited state $\ket{e}$, and a meta-stable or spin state $\ket{s}$, as shown in Figure~\ref{fig:lambda_structure}. This can be the intersystem crossing between triplet (S=1) and singlet (S=0) states or the magnetic sub-spin levels of the electron in the triplet state only (m$_s$ = +1, 0, -1).
Then, electron spins are employed for storage while the cavity plays a role in enhancing the coupling of photons. This, therefore, requires two pulses for operation: the signal pulse and the control pulse at the transition between $\ket{g} \leftrightarrow \ket{e}$ and $\ket{e} \leftrightarrow \ket{s}$, respectively.\\
\indent

\indent Characterization of the electronic transitions among a large number of defects in hBN has been also studied in~\cite{10.1002/adom.202302760}. It was unraveled that 257 defects inherit the $\Lambda$ structure in both $\Lambda$ schemes, some of which are the neutral charge, whereas others require a single charge-state manipulation to enable the feature. With the complement from a Raman-scheme decay-free quantum memory model, they could indicate the condition for achieving more than 95\% writing efficiency based on the intensity and time of a control pulse \cite{arXiv:2306.07855}. Based on the assumption that hBN is embedded in a cavity, these conditions can result in the specific decay rate of a cavity and in turn indicate the required quality factor for a cavity and its operation bandwidth. The studies revealed that, with the $\Lambda$ structure with the intersystem-crossing channel, most hBN defects require the quality factor of a cavity around 10$^6$, which is still reachable by photonics cavities \cite{pcc}.\\
\indent 
Finally, here is listed some similarities, differences, relative advantages and drawbacks of hBN-based memory compared to a well-established counterpart, NV centers in diamond, documented within the literature~\cite{10.1103/PhysRevA.89.040301,10.1364/OL.26.000361,PhysRevLett.110.213605}: First,
while NV center requires the quality factor of the cavity at 1100~\footnote{Note that its total efficiency can be enhanced by increasing the control field strength.}, hBN requires $\sim10^6$ as aforementioned. Also, the total efficiency in NV center is expected to reach 81\% together with the 99\% conditional fidelity, whereas only the writing efficiency by itself in hBN is achieved $\sim95\%$. However, these properties are incomparable owing to the different assumed $\Lambda$ structure schemes. Second, hBN defects seem to have a shorter coherence time than in NV centers. However, compatibility among other quantum systems is also vital, and this is where hBN defects likely outperform NV centers. That is, some hBN defects inherit a consistent ZPL with other quantum systems, leading to direct coupling without wavelength conversion, as demonstrated in Ref.~\cite{10.1002/adom.202302760}. In comparison to NV centers, the implementation, e.g., for the telecommunication, requires a wavelength converter to couple with another quantum system, causing some additional noises \cite{10.1103/PhysRevApplied.9.064031}.

\subsubsection{Nuclear-spin solid-state quantum memories}
Even though the framework of Raman quantum memory described in the previous section can treat a large number of defects in hBN, the prospect of exploiting nuclear spin stands out due to its high promise for longer coherence time, so that longer storage time. So far, nuclear spins in hBN have not yet been realized for the quantum memory application; however, the NV center has been demonstrated theoretically and experimentally for quantum memory \cite{10.1126/science.1139548,10.1103/PhysRevX.6.021040, 10.1038/s41534-022-00637-w,10.1103/PhysRevA.87.012301,10.1126/science.1139831,10.1038/nphys2026}. These significant achievements inspire confidence in the abilities of hBN-based quantum memory applications as well. In this quantum memory scheme, the electron and nuclear spins are utilized as the communication and storing qubits, respectively, both of which are coupled via hyperfine interaction. On the one hand, the information can be stored in nuclear spins via some nuclear spin control approach e.g. dynamic nuclear polarization with longer coherence time than electron spins. On the other hand, nuclear spins within the surrounding environment impose constraints on spin-lattice relaxation (T$_1$) and coherence time (T$_2$), as well as the storage time. These constraints are still not fully resolved, and remain as an open question to this day. \\
\indent To elaborate further, considering nuclear spins in hBN, they consist of $^{10}$B (nuclear spin (I) = 3 with 20\% natural abundance), $^{11}$B (I = 3/2 with 80\% abundance), $^{14}$N (I = 1 with 99.6\% abundance), as well as $^{15}$N (I = 1/2 with 0.4\% abundance). These non-zero nuclear spins cause considerable spin decoherence, restricting T$_1$ and T$_2$ of hBN. To tackle this issue, many simulations and experiments have been lately conducted on T$_1$ and T$_2$ enhancement as well as coherent control of nuclear spins, especially in a negative boron vacancy V$_{\text{B}}^{-1}$ defect \cite{10.1038/s41563-022-01329-8,10.1038/s41467-023-44494-3,10.1126/sciadv.abf3630,10.1021/acs.nanolett.1c02495}. 
\\
\indent As introduced in section~\ref{subsec:qsensing}, the T$_1$ time under room temperature has been reported in the range of 16 - 18 $\mu$s, independent of the magnetic field \cite{10.1038/s41467-023-44494-3,10.1126/sciadv.abf3630,10.1021/acs.nanolett.1c02495,10.1038/s41467-022-31743-0} while under low temperature, the T$_1$ is increased up to 12.5 ms \cite{10.1126/sciadv.abf3630}. For T$_2$, it depends on isotope types and yields 82 - 87 ns for natural isotopes \cite{10.1038/s41467-022-33399-2,10.1038/s41467-023-44494-3}, 46 ns for h$^{11}$BN, 62 ns for h$^{10}$BN \cite{10.1038/s41467-022-31743-0}, and 186 ns for h$^{10}$B$^{15}$N \cite{10.1038/s41467-023-44494-3}. While most prior works investigated in $^{14}$N, a recent study revealed that h$^{10}$B$^{15}$N is expected to be an optimal isotope in the sense of the longest T$_1$ and T$_2$ \cite{10.1038/s41467-023-44494-3}. This is likely because of the narrowest linewidth in the ESR spectrum of h$^{10}$B$^{15}$N. In other words, the narrow linewidth factors in less contribution from environmental spins. Note that electron spin impurities in the environment are found to contribute as another factor causing decoherence, thereby limiting T$_2$.
This was experimentally observed in the reduction of T$_1$ due to the cross relaxation \cite{10.1038/s41467-022-31743-0}.\\
\indent Compared to NV centers in a diamond, most studies exploited the nuclear spin from $^{13}$C (I = 1/2 with 1.1\% natural abundance) \cite{10.1126/science.1139548,10.1103/PhysRevX.6.021040, 10.1038/s41534-022-00637-w,10.1103/PhysRevA.87.012301,10.1126/science.1139831} and from $^{14}$N (I = 1 with 99.6\% natural abundance) \cite{10.1038/nphys2026,10.1021/acs.nanolett.7b01796} for quantum memory. Due to the low nuclear spin decoherence, the T$_1$ and T$_2$ are reported to yield 2.8 s and 1.1 s (with a single spin echo) under room temperature for $^{13}$C, respectively \cite{10.1038/s41534-022-00637-w}. These relatively long T$_1$ and T$_2$ make NV centers operational for quantum memory. For the storage process, it has been initiated by Dutt \textit{et al.} \cite{10.1126/science.1139831} that optical pumping is first applied to polarize electrons to the m$_{s}$ = 0 of the ground state. Then, the microwave (MW) pulse is applied to populate such electrons to e.g. m$_{s}$ = -1.  As such, the electron spins are now expected to interact with nearby controlled nuclear spins, hence allowing for the storage process. For the retrieval process, the reverse procedure can be applied. It should be noted that there have been later attempts to incorporate radio frequency (RF) pulse together with the microwave pulse for fast manipulation of multi-qubit registers \cite{10.1038/nature10401}. Afterward, such electron-nuclear spin coupling has been explained by the ground-state level-anti crossing, where the selection rules allow for efficient polarization transfer between the electron and nuclear spins \cite{10.1038/nphys2026,10.1103/PhysRevB.103.035307}. While NV centers show convincing suitability for quantum memory and have been developed mostly in the last decade to build a quantum network \cite{10.1063/5.0056534,10.1038/s41534-022-00637-w,10.1103/PhysRevX.6.021040,10.22331/q-2022-03-17-669},
this raises the question of whether similar storage/retrieval processes are applicable to nuclear spins in hBN.\\
\indent 
As noted above, the coherent control in hBN has been performed in $^{14}$N and $^{15}$N nuclear spins by making use of excited-state level-anti crossing and ground-state level-anti crossing based on MW and RF pulses \cite{10.1038/s41467-023-44494-3,10.1038/s41563-022-01329-8, 10.1038/s41467-022-33399-2,10.1038/s41467-023-38672-6, Ru2024}. While both are controllable, $^{15}$N can be controlled similarly to $^{13}$C thanks to the same nuclear spin, as very recently demonstrated in \cite{10.1038/s41467-023-44494-3}. 
This suggests that $^{15}$N is likely capable of performing storage/retrieval processes in principle. Nonetheless, this realization still has not been fully addressed due to the limited T$_1$ and T$_2$.  Therefore, the prolongation of T$_1$ and T$_2$ for hBN remains as one of the vital concerns for qualifying quantum memory application, thus a topic of active research. 
This study of hBN can indeed be further extended beyond V$_\text{B}^{-1}$ to explore other promising hBN defects, which poses another open area of research.
\starttocentries
\section{Further Approaches and Perspectives}
Exploring the potential connections between hBN capabilities and other techniques is an exciting possibility that, on the one hand, can result in new or improved technologies and, on the other, opens the door for addressing emerging challenges that extend beyond the current applications of hBN, as the following examples illustrate.
\begin{itemize}
\item Quantum Internet. The hBN is expected to play a crucial role in the development of future-proof quantum communication infrastructures.
Global quantum communication networks necessitate satellite-based transmission due to atmospheric and fiber channel losses. The region where the atmosphere is intensive is approximately below 10 km, and outside of this region, photons encounter negligible absorption and turbulence~\cite{Lu2022}. When looking at the big picture, hybrid systems utilizing both ground-based fiber channels and space-based free-space transmission emerge as prominent solutions for global quantum networks. This hybrid realization is already constructed as two satellites separated by a 2,600 km and 2,000 km fiber connection, using the decoy-state QKD~\cite{Chen2021}. Defects in hBN can be considered as a perfect single photon source suitable for space-based quantum communication~\cite{10.1038/s41467-019-09219-5, Ahmadi2024} as they can emit at the ideal wavelength needed for ground-to-space communication~\cite{Abasifard2024}. Such integration promises to enhance the efficiency and reliability of quantum communication networks, offering a robust solution to the challenges faced in establishing a global quantum internet. 

\item Machine Learning. The integration of machine learning techniques into the hBN domain is another decidedly active
area, which has produced accurate numerical results for characterization of hBN, including the use of concatenate neural networks~\cite{Han:2020-ML, Ramezani:2023-ML} and artificial neural networks~\cite{Masubuchi:2020-ML, Zhu:2022-ML}, as well as uncovering the effect of defects on the hBN properties~\cite{Frey:2020-ML}, but also
motivates formal developments, such as phonon dispersion relations of hBN (thermal expansion)~\cite{MORTAZAVI:2022-ML}.

\item Quantum Error Correction. 
Recent studies have demonstrated the potential of integrating hBN quantum emitters into monolithically fabricated waveguides~\cite{Gerard:2023}, which can significantly enhance the coupling efficiency of quantum emitters to photonic circuits. This integration is expected to improve by up to four times compared to conventional hybrid stacking strategies, facilitating the practical deployment of hBN in quantum photonic circuits. The ability to fabricate such devices and successfully integrate hBN single-photon emitters with waveguides opens up new possibilities for implementing topological quantum error correction codes. These codes, such as surface codes and color codes, can benefit from the stable and coherent quantum states provided by hBN defects, potentially leading to more fault-tolerant and scalable quantum computing systems.
\end{itemize}

\section{Outlook}
Quantum applications in hBN have experienced significant growth over the past decade and remain a dynamic area of research. Current hBN-based research is advancing in various directions. On one hand, a formal approach investigates the fundamental quantum properties and theoretical foundations. On the other hand, from a more practical standpoint, considerable effort is being directed toward the development of advanced methods for fabricating and characterizing hBN, a multifaceted field with key highlights discussed in the previous sections. The field is also discovering connections with seemingly unrelated topics and evolving in innovative ways. These diverse directions are poised to yield exciting outcomes in the coming years, such as identifying new defect subfamilies, enhancing the efficiency of quantum information processing, bridging the gap between fundamental research and technological applications with commercial viability.\\
\indent
In addition, mature hBN-based methodologies are well established as competitive tools for the study of solid-state quantum systems. As reviewed in this paper, these techniques not only ease the process for newcomers to try for an existing problem but also offer a platform for more experienced researchers to explore novel approaches or establish new  connections between hBN and other disciplines of study.

\section*{Notes}
The authors declare no competing financial interest.

\begin{acknowledgments}
This research is part of the Munich Quantum Valley, which is supported by the Bavarian state government with funds from the Hightech Agenda Bayern Plus. This work was funded by the Deutsche Forschungsgemeinschaft (DFG, German Research Foundation) - Projektnummer 445275953, under Germany's Excellence Strategy - EXC-2111-390814868, and as part of the CRC 1375 NOA project C2. The authors acknowledge support from the Federal Ministry of Education and Research (BMBF) under grant number 13N16292 (ATOMIQS) and by the German Space Agency DLR with funds provided by the Federal Ministry for Economic Affairs and Climate Action BMWK under grant numbers 50WM2165 (QUICK3), and 50RP2200 (QuVeKS). C.C. is grateful to the Development and Promotion of Science and Technology Talents Project (DPST) scholarship by the Royal Thai Government. This research is supported by the Australian Research Council (CE200100010, FT220100053, DP240103127) and the Office of Naval Research Global (N62909-22-1-2028). This work is supported by the Scientific and Technological Research Council of Turkey (TUBITAK) under project numbers 117F495 and 118F119. S.A. acknowledges the support by the Turkish Academy of Sciences (TUBA-GEBIP; The Young Scientist Award Program) and the Science Academy of Turkey (BAGEP; The Young Scientist Award Program).
\end{acknowledgments}

\bibliography{main}
\section*{Related Sources}
\begin{itemize}
    \item \url{https://h-bn.info} \\ Chanaprom Cholsuk, Ashkan Zand, Asli Cakan, Tobias Vogl.\\ \cite{Cholsuk2024a}{\it{The Journal of Physical Chemistry C, 2024, 128, 30, 12716.}}
    \item \url{https://cmr.fysik.dtu.dk/qpod/qpod.html} \\ Bertoldo, F., Ali, S., Manti, S., Thygesen, K. S. \\ \cite{Bertoldo2022-sg}\it{npj Comput Mater 8, 56 (2022)}
\end{itemize}

\clearpage

\clearpage

\end{document}